\documentclass{article}

\usepackage{microtype}
\usepackage{graphicx}
\usepackage{subfigure}
\usepackage{amsmath}
\usepackage{amsfonts}
\usepackage{relsize}
\usepackage{float}

\usepackage{booktabs} % for professional tables
\usepackage{hyperref}

\usepackage{soul,color}

\usepackage[accepted]{icml2023}

\icmltitlerunning{Efficient Sequence Transduction by Jointly Predicting Tokens and Durations}

\begin{document}

\twocolumn[
\icmltitle{Efficient Sequence Transduction by Jointly Predicting Tokens and Durations}

% It is OKAY to include author information, even for blind
% submissions: the style file will automatically remove it for you
% unless you've provided the [accepted] option to the icml2021
% package.

% List of affiliations: The first argument should be a (short)
% identifier you will use later to specify author affiliations
% Academic affiliations should list Department, University, City, Region, Country
% Industry affiliations should list Company, City, Region, Country

% You can specify symbols, otherwise they are numbered in order.
% Ideally, you should not use this facility. Affiliations will be numbered
% in order of appearance and this is the preferred way.
\icmlsetsymbol{equal}{*}

\begin{icmlauthorlist}
\icmlauthor{Hainan Xu}{nv}
\icmlauthor{Fei Jia}{nv}
\icmlauthor{Somshubra Majumdar}{nv}
\icmlauthor{He Huang}{nv}
\icmlauthor{Shinji Watanabe}{cmu}
\icmlauthor{Boris Ginsburg}{nv}
\end{icmlauthorlist}

\icmlaffiliation{nv}{NVIDIA, USA}
\icmlaffiliation{cmu}{Carnegie Mellon University, PA, USA}

\icmlcorrespondingauthor{Hainan Xu}{hainanx@nvidia.com}
% \icmlcorrespondingauthor{Eee Pppp}{ep@eden.co.uk}

% You may provide any keywords that you
% find helpful for describing your paper; these are used to populate
% the "keywords" metadata in the PDF but will not be shown in the document
% \icmlkeywords{speech recognition, ASR, Transducer, transducers}

 \vskip 0.3in]

% this must go after the closing bracket ] following \twocolumn[ ...

% This command actually creates the footnote in the first column
% listing the affiliations and the copyright notice.
% The command takes one argument, which is text to display at the start of the footnote.
% The \icmlEqualContribution command is conventional text for equal contribution.
% Remove it (just {}) if you do not need this facility.

\printAffiliationsAndNotice{}  % leave blank if no need to mention equal contribution
%\printAffiliationsAndNotice{\icmlEqualContribution} % otherwise use the conventional text.

% \begin{abstract}
% In this paper, we propose a novel \emph{Token-and-Duration Transducer}  (TDT) model for speech recognition.
% TDT has similar model architecture compared to conventional transducers but extends the model by jointly predicting tokens and durations of the emitted token.
% This is achieved by using the model's joint network to generate two sets of outputs, both are independently normalized to generate distributions over tokens and durations.
% During inference, instead of conventional transducer models which process the encoder output frame by frame,  the duration output of a TDT model can be used to direct the decoding process to skip frames in processing and thus
% runs faster than conventional inference.
% In order to bring more speed up, we introduce special TDT model training methods making the model favor longer durations in its outputs. 
% With experiments on four languages in speech recognition tasks, we see TDT models achieve comparable accuracy compared to conventional transducers on all datasets.
% However, TDT models allow the decoding process to skip up to 80\% of input frames, and this could bring consistent and significant inference speed-up over conventional transducers, with up to 2.20X for English Librispeech test-other, 
% 2.93X for German Multilingual Librispeech and 2.14X for Spanish Call-home.
% We open-source our implementation so that a broader research community can benefit from this method.

% \end{abstract}

\begin{abstract}
This paper introduces a novel \emph{Token-and-Duration Transducer} (TDT) architecture for sequence-to-sequence tasks.
TDT extends conventional RNN-Transducer architectures by jointly predicting both a token and its duration, i.e. the number of input frames covered by the emitted token. This is achieved by using a joint network with two outputs which are independently normalized to generate distributions over tokens and durations. 
During inference, TDT models can skip input frames guided by the predicted duration output, which makes them significantly faster than conventional Transducers which process the encoder output frame by frame.  
TDT models achieve both better accuracy and significantly faster inference than conventional Transducers on different sequence transduction tasks. 
TDT models for Speech Recognition achieve  better accuracy and up to 2.82X faster inference than  conventional Transducers.
TDT models for Speech Translation achieve an absolute gain of over 1 BLEU on the MUST-C test compared with conventional Transducers, and its inference is 2.27X faster.
In Speech Intent Classification and Slot Filling tasks,  TDT models improve the intent accuracy by up to over 1\% (absolute) over conventional Transducers, while running up to 1.28X faster.
Our implementation of the TDT model will be open-sourced with the NeMo (\url{https://github.com/NVIDIA/NeMo}) toolkit.  
\end{abstract}

\section{Introduction}

% speech recognition technology has gradually become an integral part of people's daily lives, with applications ranging from voice assistants on smartphones to voice-controlled home appliances and transcription software in the workplace. Its ability to facilitate communication and improve efficiency has made it an important tool for individuals with disabilities, as well as a valuable resource for professionals looking to save time and increase productivity.

% After decades of research, 
Over the past years, automatic speech recognition (ASR) models have undergone shifts from  conventional hybrid models 
\cite{jelinek1998statistical,woodland1994large,povey2011kaldi} to end-to-end ASR models, including attention-based encoder and decoder (AED) models
\cite{chorowski2015attention,chan2015listen}, Connectionist 
Temporal Classification
(CTC) \cite{graves2006connectionist}, 
and Transducers \cite{graves2012sequence}.
% For their superior performance and ease of deployment, they have gradually replaced conventional models and now are powering a number of speech products.
Those models are commonly used in academia and industry, and there exist  open-source toolkits with efficient implementation for those methods, including ESPNet \cite{watanabe2018espnet}, NeMo \cite{kuchaiev2019nemo}, Espresso \cite{wang2019espresso}, SpeechBrain \cite{ravanelli2021speechbrain} etc. 

This paper focuses on Transducer models.
There have been a significant number of works that improve different aspects of the original Transducer \cite{graves2012sequence}.
For example, the original LSTM encoder of transducer models has been replaced with Transformers \cite{tian2019self,yeh2019transformer,zhang2020transformer}, Contextnet \cite{contextnet} and Conformers \cite{gulati2020conformer}. 
% The encoder of transducer models has been extensively researched, with methods like Transformers\cite{yeh2019transformer,zhang2020transformer}, and Conformers \cite{gulati2020conformer} replacing original LSTM-based encoders.
Decoders for transducers are well-investigated as well, e.g. 
\cite{Ghodsi2020stateless} used stateless decoders instead LSTM decoders;
\cite{shrivastava2021echo} proposed \emph{Echo State Networks} and showed that a decoder with random parameters could perform as well as a well-trained decoder.
The loss function of Transducers has also been an active research area. 
FastEmit  \cite{yu2021fastemit} introduces biases in the gradient update of the transducer loss to reduce latency. Multi-blank Transducers \cite{xu2022multi} introduce a generalized Transducer architecture and loss function with \emph{big blank} symbols that cover multiple frames of the input. %, which brings both inference speed-up and accuracy gains.

RNN-Ts have achieved impressive accuracy in speech tasks, but
% while its training requires significant computational resources.
% , both for training and inference.
% While the training of those models is a one-time cost that can be amortized, 
the auto-regressive decoding makes their inference computationally costly. %, especially in limited-resource settings, e.g. for mobile devices. 
% In this paper, we propose an extension of the Transducer
To alleviate this issue, we propose a new Transducer architecture that jointly predicts  a token and its duration. The predicted token duration 
% output of the model 
can direct the model decoding algorithm to skip frames during inference.  We call it a TDT (Token-and-Duration Transducer) model.
% This paper is organized as follows. In Section \ref{background}, we give a brief introduction to 
% Transducer models. In Section \ref{tdt},
% we describe our novel TDT Transducer model.
% We report our experiment results in Section \ref{results}. We present some interesting analyses in Section \ref{analysis} and lastly, we conclude the paper and discuss future work in Section \ref{future}.
The primary contributions of this paper are:
\begin{enumerate}
    \item A novel Token-and-Duration Transducer (TDT) architecture that jointly predicts a token and its duration.
    \item An  extension of the forward-backward algorithm  to derive the analytical solution of
    % of the gradient for TDT models. In particular, we derive 
    the gradients of the TDT model. We also derive gradients of pre-softmax logits for the token prediction inspired by \emph{Transducer function-merging} \cite{li2019improving}. 
    \item TDT models achieve better accuracy and significant inference speed-up compared to original RNN-Ts for 3 different tasks -- speech recognition, speech translation,  and spoken language understanding.
    \item TDT-based ASR  models are more robust to noise than conventional RNN-Ts models, and  they don't suffer from the performance degradation for speech corresponding to the text with repeated tokens.
    % TDT models achieve new state-of-the-art for ASR on Voxpopuli and for SLU.
\end{enumerate}
Our TDT model implementation will be open-sourced with NVIDIA's NeMo \footnote{\url{https://github.com/NVIDIA/NeMo/}. Pull request \url{https://github.com/NVIDIA/NeMo/pull/6536}} toolkit.

\section{Background: Transducers} \label{background}
%\subsection{conventional Transducers}

% \begin{figure}[h]
%     \centering
%     \includegraphics[scale=0.35]{arch.png}
%     \caption{old arch [todo]}
%     \label{old_arch}
% \end{figure}

% \begin{figure}[h]
%     \centering
%     \includegraphics[scale=0.25]{one_ori.png}
%     \caption{origin node action [todo] }
%     \label{old_lattice}
% \end{figure}

\begin{figure}
    \centering
    \includegraphics[scale=0.078]{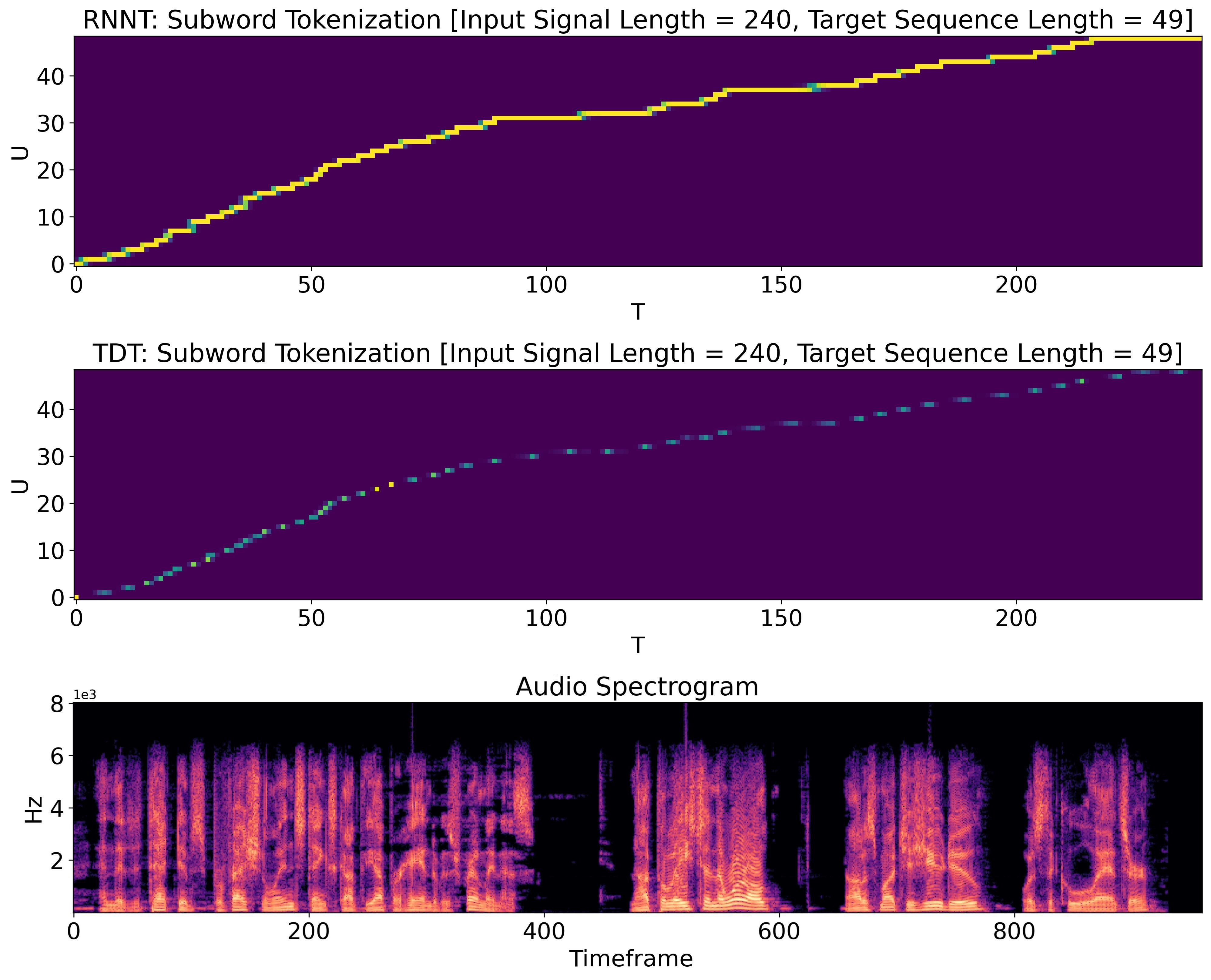}
    \caption{From top to bottom: alignments generated with conventional RNNT, TDT models with config  [0-8], and the corresponding spectrogram. Each unit in the $T$ axis of the alignment corresponds to 4 frames in the spectrogram due to subsampling. Note, TDT model learns to skip frames. Long skips are not frequently used in the audio where speech is present, but for the 4 relatively silent segments in the audio, where conventional RNN-T's alignment shows mostly horizontal lines, the TDT model uses long durations to skip the majority of frames. }
    \label{first_figure}
\end{figure}

An RNN-Transducer\footnote{When originally proposed, an RNN-Transducer uses a recurrent network as the encoder hence the name \emph{RNN-T};  nowadays Transducers usually adopt more sophisticated networks involving self-attention in their encoder. In this paper, we use the words \emph{RNN-T} and \emph{Transducer} interchangeably to represent any encoder-decoder-joiner model that uses the Transducer loss.} \cite{graves2012sequence} consists of an  encoder, 
% (also referred to as an acoustic encoder, or a transcription network), 
a decoder (or a prediction network), and a joint network (or a joiner). 
The encoder and decoder extract higher-level representations of the acoustic and text  and feed the output to the joint network, which generates a probability distribution over the vocabulary.
The vocabulary includes a special blank symbol \O; a text sequence could be augmented by adding an arbitrary number of blanks between any adjacent tokens. During training, we maximize the log-probability $\log P_\text{RNNT}(y | x)$ for an audio utterance $x$ with corresponding text $y$, which requires summing over all possible ways to augment the text sequence to match the audio:% 
\begin{equation}
\begin{aligned}
\mathcal{L}_\text{RNNT} (y|x) =  &  \log P_\text{RNNT}(y | x) \\
= & \log \mathlarger\sum_{\pi: B^{-1}(\pi) = y} P_\text{frame-level}(\pi | x),
\end{aligned}
\end{equation}
where $\pi$ represents an augmented sequence (including \O), 
$B(.)$ is the operation to augment a sequence by adding blanks, and $B^{-1}$ is the inverse of the operation $B$, which  
% . In this context, the $A^{-1}$ operation 
removes all the blanks in the sequence. 

Computing  $P_\text{RNNT} (y|x)$ using its definition is intractable since it needs to sum over exponentially many possible augmented sequences. In practice, the probability can be efficiently computed with the forward variables $\alpha(t, u)$ or backward variables $\beta(t, u)$, which are calculated recursively:
\begin{align}
 \begin{split}\label{alpha}
    \alpha(t, u)  &=  \alpha(t - 1, u) P(\text{\O}|t-1, u) \\
     & +  \alpha(t, u - 1) P(y_u|t, u - 1)
  \end{split} \\
  \begin{split}\label{beta}
    \beta(t, u) & =  \beta(t + 1, u) P(\text{\O}|t, u) \\
    & + \beta(t, u + 1) P(y_{u+1}|t, u).
\end{split}
\end{align}
with recursion base conditions $\alpha(1,0) = 1$ and $\beta(T, U) = P(\text{\O}| T, U)$.
%
%
%  \begin{equation}\label{alpha}
 % \begin{split}
 %    \alpha(t, u)  &=  \alpha(t - 1, u) P(\O|t-1, u) \\
 %     & +  \alpha(t, u - 1) P(y_u|t, u - 1)
 %  \end{split} \\
%  \end{equation}
%  \begin{equation}\label{beta}
%   \begin{split}
%     \beta(t, u) & =  \beta(t + 1, u) P(\O|t, u) \\
%     & + \beta(t, u + 1) P(y_{u+1}|t, u).
% \end{split}
%  \end{equation}
In order to make this recursion well-defined, we require that  both $\alpha(t, u)$ and $\beta(t, u)$  are zero outside domain $1 \leq t \leq T$  and $0 \leq u \leq U$.
 % Any $\alpha$ and $\beta$ variables outside of the condition . 
% in the computation. With the $\alpha$ and $\beta$ variables defined,
With those quantities defined,
$P_\text{RNNT}(y|x)$ could be computed with either the $\alpha$ or $\beta$ efficiently: 
% We have
\begin{align*}
    % & P(y | x) = \alpha(T, U) \O(T, U) \\
    & P_\text{RNNT}(y | x) = \alpha(T, U) P(\text{\O} |T, U) = \beta(1, 0).
    % & P(y | x) = \beta(1, 0).
\end{align*}

Then we could compute the loss function as,
\begin{equation}
    \mathcal{L}_\text{RNNT}(y | x) = \log P_\text{RNNT}(y | x).
\end{equation}

%
% \begin{equation}
%     P(y | x) = \alpha(T, U) \O(T, U)
% \end{equation}
% and
% \begin{equation}
%     P(y | x) = \beta(1, 0).
% \end{equation}

% We point out here that although only one of $\alpha$ and $\beta$ variables is enough to compute the loss efficiently, both are needed in order to efficiently compute its gradient, which will be addressed in more details in Section \ref{gradient}.
 
 \section{Token-and-Duration Transducers}\label{tdt}
 
 \begin{figure}[t]
    \centering
    \includegraphics[scale=0.35]{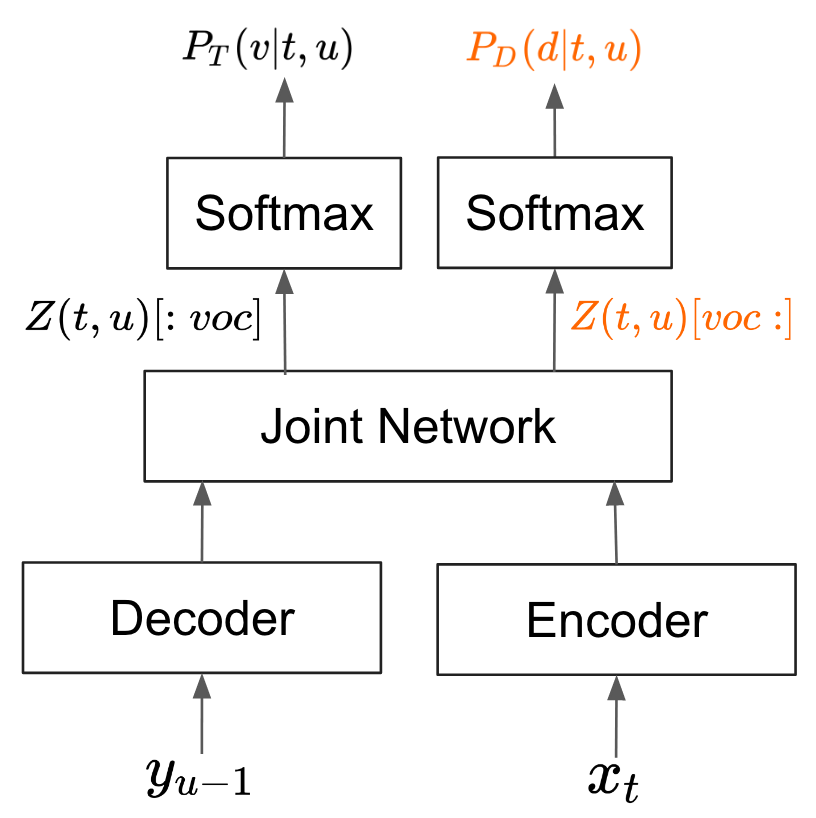}
    \caption{Architecture of a TDT model, which contains an encoder, a decoder, and a joint network. The TDT joint network emits two sets of output, one for the output token $Z(t,u)[\text{:voc}]$, and the other for the duration of the token  $Z(t,u)[\text{voc:}]$. The two distributions are jointly trained during model training.}
    \label{new_arch}
\end{figure}

\begin{figure}[t]
    \centering
    \includegraphics[scale=0.3]{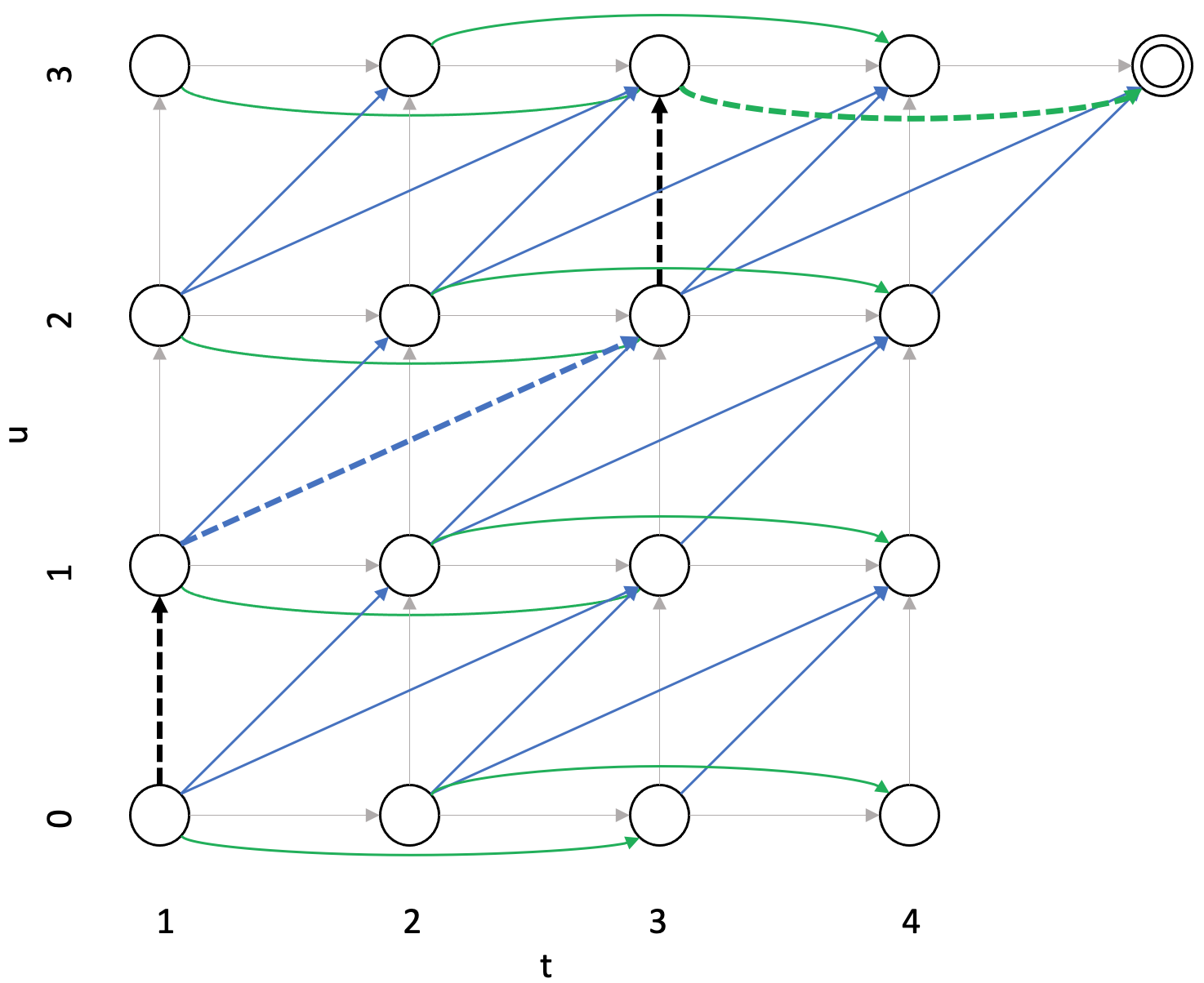}
    \caption{Output probability lattice of  TDT model with supported durations \{0,1,2\}. We follow the convention in \cite{graves2012sequence} making $t$ start with 1 and $u$ with 0. The probability of observing the first $u$ output labels  in the first $t$  frames is represented by node $(t, u)$.  Dotted arrows constitute a complete path in the lattice. }   
     % {\color{our_green} Green} arcs here denote big blanks with duration $2$, represented as $\O_2(t,u)$ and {\color{our_blue} blue}  denotes big blanks with duration $3$, represented as $\O_3(t,u)$. The dashed line is a possible path with the proposed method.}
    \label{new_lattice}
\end{figure} 
A Token-and-Duration Transducer (TDT) differs from conventional transducers in that it  predicts the token duration of the current emission. 
Namely, the TDT joiner  generates two sets of output, one for the output token, and the other for the duration of the token (see Fig.~\ref{new_arch}).
\footnote{In our implementation, the two outputs are disjoint sub-vectors of the joiner output. For example, let's take vocabulary size (voc) of 1024 (including $\O$) that supports durations \{0,1,2,3,4\} (a total of 5 durations). The last layer of the joiner maps the hidden  activation to a tensor \url{joiner_out} of size 1024 + 5 = 1029. Then \url{joiner_out[:1024]} and \url{joiner_out[1024:]} are independently normalized  to generate the two sets of distributions.}
% To illustrate the mathematical details of TDT models, 
Let us first define a joint probability $P(v, d | t, u)$ as the probability of generating token $v$ ($v$ could either be a text token or $\O$), with duration $d$ at location $(t, u)$.
We assume that token and durations are conditionally independent:  
% and compute $P(v, d | t, u)$ as,
\begin{equation}
    P(v, d | t, u) =        P_T(v | t, u) P_D(d | t, u)
\end{equation}
where $P_T(.)$ and $P_D(.)$ correspond to the token distribution and duration distribution, respectively.
Next, we can compute the forward variables $\alpha(t,u)$:
\begin{equation}
\label{eqn:alphas}
\begin{split}
    \alpha(t, u)  &=  \mathlarger{\sum}_{d\in \mathcal{D}\setminus{\{0\}}}\alpha(t - d, u) P(\O, d | t-d, u)  \\
    & +\mathlarger\sum_{d\in \mathcal{D}} \alpha(t - d, u - 1)  P(y_u, d | t-d, u-1)
\end{split}
\end{equation}
with the same base condition $\alpha(1,0) = 1$ as that of the conventional Transducer. 
Note, this Equation differs from \ref{alpha} in that, for both non-blank  and blank emissions, we need to sum over  durations in $\mathcal{D}$ to consider all possible contributions from  states that can reach $(t, u)$, weighted by the corresponding duration probabilities.\footnote{We disallow blank emission with duration 0, thus the sum is over $\mathcal{D} \setminus \{0\}$ for the blank emission. This makes the model not strictly probabilistic unless we renormalize the duration probabilities excluding duration = 0 for blank emissions computation. Although in practice we find that this does not matter, since duration=0 is in general rarely predicted according to Figure \ref{fig:distribution}, and this design makes the derivation of gradients much easier.}
Readers are encouraged to compare those Equations with the transition arcs in Figure \ref{new_lattice} to see the connections.
% Original equation (Hainan)
The total output probability $P_\text{TDT}(y|x)$ is computed through $\alpha$ at the terminal node:
\footnote{This equation, and the base condition of $\beta$ are slightly different from the common definition used for conventional RNNT, although they are equivalent to the standard definition. For TDT though, this notation will make the boundary case much easier. In the recursion, we have $\beta(T + 1, u) = \alpha(T + 1, u) = 0, \forall u \neq U$.}
\begin{equation}
    P_\text{TDT}(y | x) = \alpha(T + 1, U)
\end{equation}
The backward variables ($\beta$) are computed as,
\begin{equation}
\label{eqn:betas}
\begin{split}
    \beta(t, u) & =  \mathlarger\sum_{d\in \mathcal{D}\setminus{\{0\}}} \beta(t + d, u) P(\O, d | t, u)\\
    & + \mathlarger\sum_{d\in \mathcal{D}} \beta(t+d, u + 1) P(y_{u+1}, d | t, u)\\
    % & + \begin{cases}
    %     P(\O, d|t, u), & \exists d\in \mathcal{D}: d>1, t+d = T + 1 \\
    %     0, & \text{otherwise} \\
    % \end{cases}
\end{split}
\end{equation}
with the base condition $ \beta(T + 1, U) = 1$. 
% \begin{equation*}
%     \beta(T, U) = P(\O, 1| T, U).
% \end{equation*}
The probability of the whole sequence is $ P_\text{TDT}(y | x) = \beta(1,0)$. 
% \begin{equation}
%     P(y | x) = \beta(1,0)
% \end{equation}

With those quantities defined, we define TDT loss as
\begin{align}
    \label{eqn:TDT_loss}
    \mathcal{L}_\text{TDT} &= -\log P_\text{TDT}(y | x)
    %&= -\log \left( P_T(y | x) + P_D(y | x) \right) 
\end{align}
%The output probability lattice of TDT is shown in Fig.~\ref{new_lattice} .

\subsection{TDT Gradient Computation} \label{gradient}

We derive an analytical solution for the gradient of the TDT loss, since automatic differentiation  for transducer loss is highly inefficient.
\footnote{The detailed derivation for all gradients is in Appendix A.} % Instead, we manually compute the gradients. 
The gradient of the TDT loss $\mathcal{L}$ has two parts. 
The first part is the gradient with respect to the token probabilities $P_T(v | t, u)$: 
% are shown in Eq.~\ref{token}: 
\begin{equation} 
    \label{token}
    \frac{\partial \mathcal{L}_\text{TDT}}{\partial P_T(v|t, u)} = -\frac{\alpha(t,u) b(v, t, u) }{P_\text{TDT}(y|x)} 
\end{equation}
where $\alpha(t,u)$ are defined in Equation \ref{eqn:alphas} and $b(v, t, u)$ is define as:
\begin{equation} \label{eqn:B_vtu}
    b(v, t, u) = \begin{cases}
        \mathlarger\sum\limits_{d\in \mathcal{D}} \beta(t+d,u+1) P_D(d | t, u),  & v = y_{u+1} \\
        \mathlarger\sum\limits_{d\in \mathcal{D} \setminus{\{0\} }} \beta(t+d,u) P_D(d| t, u),  & v = \O \\
        0,             & \text{otherwise.} \\
    \end{cases}
\end{equation}

Note, $b(v,t,u) $ can be interpreted as a weighted sum of  $\beta$'s that are reachable from $(t, u)$, where the weights are from the duration probabilities.

The second part is the  gradient with respect to the duration probabilities $P_D(d | t, u)$:
 % are shown in Eq.~\ref{duration}:
\begin{equation}
    \label{duration}
    \frac{\partial \mathcal{L}_\text{TDT}} {\partial P_D(d|t,u)} = -\frac{\alpha(t, u) c(d, t, u)}{P(y|x)}
\end{equation}
where
\begin{equation} \label{eqn:C_vtu}
    c(d, t, u) = \begin{cases}
        \beta(t, u + 1) P_T(y_{u+1} | t, u), & d = 0 \\
        \beta(t + d, u + 1) P_T(y_{u+1} | t, u)  \\
       \ \ \ + \beta(t + d, u) P_T(\O | t, u), & d > 0. \\
    \end{cases}
\end{equation}

\subsection{Gradient  with Transducer  Function Merging}
The $P_T(v|t, u)$ terms in the Transducer loss are usually computed with a softmax function. Thus the gradients of the TDT loss have to go through the gradient of the softmax function to be passed to the previous layers, which could be costly.
 % proposed. 
% Unlike the Equations given in the original Transducer paper from \cite{graves2012sequence} which computes
% the gradient with respect to $P(k | t, u)$, \cite{li2019improving}
We use \emph{Transducer function merging}  proposed in \cite{li2019improving} to directly compute the gradient of the Transducer loss with respect to the \emph{pre-softmax} logits ($h^v(t,u)$):
% The gradient to pre-softmax token logits, shown in Equations \ref{rnnt_grad1} and \ref{functio_merging_tdt},
\begin{equation} \label{rnnt_grad1}
\frac{\partial \mathcal{L}_\text{TDT}(y|x)}{\partial h^v(t,u)} = \frac{P_T(v|t, u)\alpha(t,u) \Big[\beta(t, u) - b(v, t, u)\Big] }{P_\text{TDT}(y | x)} 
\end{equation}
where $b(v,t,u) $ is defined in Eq.~\ref{eqn:B_vtu}.
% \begin{equation}
%     \label{functio_merging_tdt}
%      d(v, t, u) = \begin{cases}
% \mathlarger\sum\limits_{d\in \mathcal{D}} \beta(t+d, u+1) P_D(d|t,u), & v = y_{u+1} \\
% \mathlarger\sum\limits_{d\in \mathcal{D} \setminus \{0\}} \beta(t+d, u) P_D(d|t,u), & v = \O \\
% 0, & \text{otherwise.}
% \end{cases}
% \end{equation}
Note we apply function merging only to the \emph{token logits}, not \emph{duration logits} since the latter  usually has very small dimensions, and the negligible efficiency improvements do not outweigh the added complexity in implementation.

\subsection{Logits Under-normalization} \label{sec:under_norm}
We adopt the \emph{logit under-normalization} method from \cite{xu2022multi} during the training of  TDT models, in order to encourage longer durations. 
In our TDT implementations, we compute $P_T(v|t, u)$  in the log domain in order to have better numerical stability. The log probabilities $\log P_T(v|t, u)$ are computed from the logits  $h^v(t,u)$ corresponding to token $v$:
\begin{equation}
    \log P_T(v|t, u) =
    % \text{torch.nn.log\_softmax(logits)}.
    \text{log\_softmax}_{v'} (h^{v'}(t, u)).
\end{equation}
The TDT model uses the pseudo ``probability'' $P'_T(v|t,u)$  in its forward and backward computation, which \emph{under-normalize} the logits in the following way:
\begin{equation}
\label{logits}
\log P'_T(v|t, u) =
    % \text{torch.nn.log\_softmax(logits)}.
    \text{log\_softmax}_{v'} (h^{v'}(t, u)) - \sigma.
\end{equation}
%We set $\sigma = 0.05$ following in~\cite{xu2022multi}, where under-normalization was used for training multi-blank Transducers. 
The under-normalization is only used in training, which encourages TDT models to prioritize emissions of any token (blank or non-blank) with longer durations. 
The gradients that incorporate the logit under-normalization method are shown in Eq.~\ref{final}, 
\begin{equation}
\label{final}
\frac{\partial \mathcal{L}_\text{TDT}(y|x)}{\partial h^v(t,u)} = \frac{P_T(v|t, u) \alpha(t,u) \Big[\beta(t, u) - \frac{b(v, t, u)}{\exp(\sigma)}\Big] }{\exp\Big[\mathcal{L}_\text{TDT}(y|x)\Big]} 
\end{equation}
where  $b(v, t, u)$ are defined in Eq.~\ref{eqn:B_vtu}.\footnote{$P_T(.)$  in  
Eq.~\ref{final}, represents ``real'' probability, while the loss function $\mathcal{L}$ is computed with pseudo-probabilities.}
Note that Eq. \ref{final} is similar to 
Eq.~\ref{rnnt_grad1}, with the only difference being for TDT, the $b(v, t, u)$ term is scaled by $\frac{1}{\exp(\sigma)}$. %as the result of logit under-normalization.
% where
% \begin{equation}
%     \label{functio_merging_tdt_sigma}
%      b(v, t, u) = \begin{cases}
% \mathlarger\sum\limits_{d\in \mathcal{D}} \beta(t+d, u+1) P_D(d|t,u), & v = y_{u+1} \\
% \mathlarger\sum\limits_{d\in \mathcal{D} \setminus \{0\}} \beta(t+d, u) P_D(d|t,u), & v = \O \\
% 0, & \text{otherwise.}
% \end{cases}
% \end{equation}

\subsection{TDT Inference}
We compare the inference algorithms of conventional Transducer models (Algorithm \ref{RNNT_algo}) and TDT models (Algorithm \ref{TDT_algo}), which fully utilize the duration output.
%\footnote{This is  to demonstrate the interactions between the decoding procedure and the duration output of the model, and that TDT could be adopted in beam search decoding as well.} 
Note, that for TDT, an additional distribution over durations is computed from the joiner (line 9).  This duration can increment $t$ by more than one (line 13), compared with line 12 of conventional Transducer algorithm, where $t$ could only be incremented by 1 at a time, and this only happens for blank emissions. This is the key place that makes TDT inference faster.
%Note, for simplicity, we assume the ``argmax'' operation  directly returns the text token/duration length, depending on the context in the Algorithms.
\begin{algorithm}[tb]
   \caption{Greedy Inference of Conventional Transducer}
   \label{RNNT_algo}
\begin{algorithmic}[1]
   \STATE {\bfseries input:} acoustic input $x$
    \STATE enc = encoder($x$)
    \STATE hyp = []
    \STATE $t$ = 0

    \WHILE{$t <$ len(enc)}
    \STATE dec = decoder(hyp)
    \STATE joined = joint(enc[$t$], dec)
%    \STATE token\_probs = softmax(joined)
    \STATE idx = argmax(joined)
    \IF{token is not blank}
    \STATE hyp.append(idx2token[idx])
    \ELSE
    \STATE $t$ += 1 
    \ENDIF
    \ENDWHILE
    \STATE {\bfseries return} hyp 
\end{algorithmic}
\end{algorithm}

\begin{algorithm}[t]
   \caption{Greedy Inference of TDT Models}
   \label{TDT_algo}
\begin{algorithmic}[1]
   \STATE {\bfseries input:} acoustic input $x$
    \STATE enc = encoder($x$)
    \STATE hyp = []
    \STATE $t$ = 0

    \WHILE{$t <$ len(enc)}
    \STATE dec = decoder(hyp)
    \STATE joined = joint(enc[$t$], dec)
%    \STATE token\_probs = softmax(joined[:vocab\_size])
%    \STATE duration\_probs = softmax(joined[vocab\_size:])
    \STATE idx = argmax(joined[:vocab\_size])
    \STATE duration\_idx = argmax(joined[vocab\_size:])
    \IF{token is not blank}
    \STATE hyp.append(idx2token[idx])
    \ENDIF
    \STATE $t$ += duration\_idx2duration[duration\_idx]
    \ENDWHILE
    \STATE {\bfseries return} hyp 
\end{algorithmic}
\end{algorithm}

\section{Experiments}\label{results}
We evaluate our model in three different tasks: speech recognition, speech translation, and spoken language understanding.
We use the NeMo \cite{kuchaiev2019nemo} toolkit for all experiments. Unless  specified otherwise, we use \emph{Conformer-Large} for all tasks. \footnote{\url{examples/asr/conf/conformer/conformer_transducer_bpe.yaml} in \url{https://github.com/NVIDIA/NeMo}} 
For acoustic feature extraction, we use audio frames of 10 ms and window sizes of 25 ms.
Our model has a conformer encoder with 17 layers with num-heads = 8, and relative position embeddings. The hidden dimension of all the conformer layers is set to 512, and for the feed-forward layers in the conformer, an expansion factor of 4 is used. The convolution layers use a kernel size of 31. At the beginning of the encoder,  convolution-based subs-ampling is performed  with subsampling rate 4.
All models have around 120M parameters, The exact number of parameters may vary, depending on the size of the subword vocabulary and durations used with TDT models.
We use different subword-based tokenizers  for different models, which will be described in their respective sections. 
Unless specified otherwise, logit under-normalization is used during training with $\sigma = 0.05$.
For all experiments, we train our models for no more than 200 epochs, and run checkpoint-averaging performed on 5 checkpoints with the best performance on validation data, to generate the model for evaluation. We run non-batched greedy search inference \footnote{Beam search for TDT models is highly complex since the search space spans both token and duration dimensions. That being said, it is possible to come up with different pruning methods to step up the TDT beam search, which will be our future work.}  for all evaluations reported in this Section. TDT Batched inference is discussed in Section \ref{batch}. No external LM is used in any of our experiments.

\subsection{Speech Recognition}
We evaluate TDT for  English, German, and Spanish ASR. 
All ASR models uses Conformer-Large encoder with stateless decoders \cite{Ghodsi2020stateless}, which  concatenates the embeddings of the last 2 history words as
the decoder output. 
All models use Byte-Pair Encoding (BPE) \cite{sennrich2015neural} as the text representation with vocabulary size = 1024.  Fast-emit \cite{yu2021fastemit} regularization is used in all our models, with $\lambda = 0.01$.

For each language, the baseline is the conventional Transducer model. We test  TDT models with % a number of 
different $\mathcal{D}$ configurations. We choose consecutive integers as our configurations, and use a shorthand notation to represent durations from 0 to the maximum value, e.g. ``0-4'' means $\mathcal{D} = \{0,1,2,3,4\}$. All  models have been trained only on public datasets to make the experiments reproducible. 

%\subsection{Training Speed}

\subsubsection{English ASR}
Our English ASR models are trained on the Librispeech \cite{panayotov2015librispeech} set with 960 hours of speech. Speed perturbation with factors (0.9, 1.0, 1.1) is performed to augment the dataset. TDT models achieve similar  accuracy compared to the baseline (RNNT). TDT models are also  significantly faster in inference, up to 2.19X and 2.12X with config 0-8 for test-clean and test-other, respectively (see Tables \ref{tab:librispeech_clean} and \ref{tab:librispeech}).

\begin{table}
    \centering
    \begin{tabular}{cccc}
    \toprule
         TDT config & WER(\%) & time(s) & rel. speed-up \\
    \midrule
    RNNT  & 2.14 & 256 & - \\
        0-2 & 2.35 & 175 & 1.46X \\
        0-4 & 2.17 & 129 & 1.98X \\
        0-6 & 2.14 & 119 & 2.15X \\
        0-8 & 2.11 & 117 & 2.19X \\
    \bottomrule
    \end{tabular}
    \caption{English ASR, Librispeech test-clean. TDT vs RNNT: WER, decoding time, and relative speed-up against the RNNT.
    }
    \label{tab:librispeech_clean}
\end{table}

\begin{table}
    \centering
    \begin{tabular}{cccc}
    \toprule
        TDT config & WER(\%) & time(s) & rel. speed-up \\
    \midrule
   RNNT & 5.11 & 244 & - \\
   \hline
        0-2 & 5.50 & 171 & 1.43X \\
        0-4 & 5.06 & 128 & 1.91X \\
        0-6 & 5.05 & 118 & 2.07X \\
        0-8 & 5.16 & 115 & 2.12X \\
    \bottomrule
    \end{tabular}
    \caption{English ASR, Librispeech test-other. TDT vs RNNT: WER, decoding time, and relative speed-up against the RNNT. 
    % Relative speed-up is measured against the baseline.
    }
    \label{tab:librispeech}
\end{table}

\begin{table}[t]
    \centering
    \begin{tabular}{cccc}
    \toprule
        TDT config & WER(\%) & time(s) & rel. speed-up  \\
    \midrule
    RNNT & 19.84 & 47  & - \\
       \hline
        0-2  & 17.95 & 33 & 1.42X\\
        0-4  & 18.57 & 26 & 1.81X\\
        0-6  & 18.06 & 24 & 1.96X\\
        0-8  & 18.73 & 24 & 1.96X\\
    \bottomrule
    \end{tabular}
    \caption{Spanish ASR, on CallHome dataset. TDT vs RNNT: WER, decoding time, and relative speed-up against RNNT. }
    \label{tab:spanish}
\end{table}

\iffalse
{
\begin{table}
    \centering
    \begin{tabular}{cccc}
    \toprule
        TDT config & WER(\%) & time(s) & rel. speed-up  \\
    \midrule
        RNNT & 8.03 & 231 & - \\
           \hline
        0-2      & 7.78 & 160 & 1.44X \\
        0-4      & 7.66 & 123  & 1.88X \\
        0-6      & 7.66 & 114  & 2.03X \\
        0-8      & 7.49 & 112 & 2.06X \\

    \bottomrule
    \end{tabular}
    \caption{German ASR,  Voxpopuli. TDT vs RNNT: WER(\%), decoding time(s), and relative speed-up for different TDT configs. Relative speed-up is measured against the RNNT.
    % Relative speed-up is measured against the baseline.
    }
    \label{tab:german_vox}
\end{table}
%
}\fi 
%
\begin{table}[h!]
    \centering
    \begin{tabular}{cccc}
    \toprule
        TDT config & WER(\%) & time(s) & rel. speed-up  \\
    \midrule
        RNNT   & 3.99 & 558  & - \\
        \hline
        0-2 & 4.10 & 352 & 1.59X \\
        0-4 & 3.93 & 232 & 2.41X \\
        0-6 & 4.00 & 207 & 2.70X \\
        0-8 & 3.95 & 198 & 2.82X \\
    \bottomrule
    \end{tabular}
    \caption{German ASR on MLS set. TDT vs RNNT: WER, decoding time, and relative speed-up against RNNT. 
    % Relative speed-up is measured against the baseline.
    }
    \label{tab:german_mls}
\end{table}
\subsubsection{Spanish ASR}
Our Spanish models are trained on combination of Mozilla Common Voice (MCV) \cite{ardila2019common}, Multilingual Librispeech (MLS) \cite{pratap2020mls}, Voxpopuli \cite{wang2021voxpopuli}, and Fisher {(LDC2010S01)} 
%Callhome (LDC96S35), TEDx \cite{mena_2019} and Mediaspeech  \cite{mediaspeech2021} 
dataset with 1340 hours in total. We evaluate our model on the Spanish Callhome (LDC96S35) test set. We see consistent WER improvement with TDT models compared to RNNT, with up to almost 2 absolute WER points for 0-2 TDT, and over 1 absolute WER point for our fastest model with configuration = 0-8 (See  Table \ref{tab:spanish}). 
TDT models are much faster than RNNT, with maximum speed up factor of 1.96 for  0-6 and 0-8 TDT configurations.

\subsubsection{German ASR}
 The German ASR was trained on MCV, MLS, and Voxpopuli datasets, with a total of around 2000 hours. Models are evaluated on
 % Voxpopuli (Table \ref{tab:german_vox} ) and 
 MLS test set. 
TDT models have accuracy similar or better than RNNTs (Table \ref{tab:german_mls}). We also observe 2.82X speed up on German MLS test for TDT 0-8 configuration, which is higher than on other datasets.\footnote{The large speed-up is related to the fact that MLS dataset has longer text: for other datasets (Librispeech, Spanish Callhome), an utterance contains on average between 20 and 40 subword tokens, but MLS has 68. While the encoder  is easily parallelized, the decoding is autoregressive and it has to be performed sequentially. Therefore the model spends more time in the decoding search, so we see a larger speed up.}

\subsection{Speech Translation}

We evaluate TDT models on English-to-German Speech Translation. 
% \footnote{The goal of ST experiments is not to catch up with SOTA, but only to compare TDT with conventional Transducers on this task. 
% % Note, that the monotonic nature of Transducers maybe not an ideal match for translation tasks that require reordering.
% }
For baseline, we directly applies a Conformer Transducer model on speech translation datasets, without any changes to the model. To the best of our knowledge, at the time of writing this paper, there are no reported results with such models, while the closest are from \cite{xue2022large}, where the authors added attention pooling to the joint network in the Transducer model in order to better model reordering in translations.
We train our models on a combination of MUST-C V2  \cite{cattoni2021must}, CoVoST V2 \cite{wang2020covost}, ST-TED \cite{niehues-etal-2018-iwslt}, Europarl-ST \cite{iranzo2020europarl}, as well as English audio data from CommonVoice v6 and VoxPopuli v2 with German text generated with an NMT model trained on WMT21 \cite{farhad2021findings} data.
Tokenization of the text uses the YouTokenToMe \footnote{\url{https://github.com/VKCOM/YouTokenToMe}} tokenizer, with a 16k vocabulary size.
All models are trained from scratch using the aforementioned training set. 
% and no pre-trained model is used, and not trained with multilingual datasets. 
 Table \ref{ST_results}
shows our results on the MUST-C V2 Test dataset. For reference, we also include results of the best publicly available model at the time of writing  from \cite{indurthi2021task}.\footnote{\url{https://paperswithcode.com/sota/speech-to-text-translation-on-must-c-en-de}.} Note,  the models are not directly comparable with our RNNT and TDT models, since they are trained on different datasets. We see that while baseline RNNT gives a decent result of a BLEU score of 23.21, TDT models consistently improve that with up to 1.26 BLEU score points, and the inference is up to 2.27X faster (see  Table \ref{ST_results}). TDT models demonstrate a stronger modeling capacity over conventional RNNTs.

\begin{table}[h]
    \centering
    \begin{tabular}{cccc}
    \toprule
        Model & BLEU (\%) & time(s) & speed-up \\
    \midrule
\cite{indurthi2021task}          & 28.88 & N/A & N/A  \\
%         \cite{le2021lightweight} & 24.63 & - & - \\
    \midrule
        RNNT & 23.21 & 218  &  - \\
        TDT 0-2      & 24.03 & 143  & 1.52X \\
        TDT 0-4      & 24.15 & 106  & 2.06X \\
        TDT 0-8      & 24.47 & 96   & 2.27X \\
        \bottomrule
    \end{tabular}
    \caption{Speech Translation, MUST-C V2 Test dataset. TDT vs RNNT: BLEU score, inference time, and relative inference speed-up of different speech translation models.}
    \label{ST_results}
\end{table}

\begin{table}[t]
\resizebox{\columnwidth}{!}{
\begin{tabular}{ccccc}
\toprule
Model & \begin{tabular}[c]{@{}c@{}}\#params\\ (M)\end{tabular} & \begin{tabular}[c]{@{}c@{}}intent\\ acc.\end{tabular} & \begin{tabular}[c]{@{}c@{}}SLURP \\ F1\end{tabular} & \begin{tabular}[c]{@{}c@{}}rel. \\ speedup\end{tabular} \\ \hline
SpeechBrain & 96 & 87.7 & 76.19 & N/A \\
ESPnet-SLU & 109 & 86.52 & 76.91 & N/A \\
\midrule
RNNT & 119 & 88.53 & 79.41 & - \\
TDT 0-2 & 119 & 87.12 & 79.43 & 1.17x \\
TDT 0-4 & 119 & 89.85 & 80.03 & 1.17x \\
TDT 0-6 & 119 & 89.28 & 80.61 & 1.28x \\
TDT 0-8 & 119 & 90.07 & 79.90 & 1.28x \\ 
\bottomrule
\end{tabular}
}
\caption{Speech intent classification and slot filling on SLURP dataset. TDT vs RNNT: Relative speed-up against RNNT.}
\label{tab:slu}
\end{table}

\subsection{Spoken Language Understanding}

In this section, we apply TDT models to spoken language understanding (SLU), specifically the \emph{Speech Intent Classification and Slot Filling} (SICSF) task, which takes audio as input to detect user intents and extract the corresponding lexical fillers for detected entity slots ~\cite{slurp}. An intent is composed of a scenario type and an action type, while slots and fillers are represented by key-value pairs. The ground-truth intents and slots of input are organized as a Python dictionary, represented as a Python string. The SICSF task is to predict this text based on the input audio. Experiments are conducted using the SLURP~\cite{slurp} dataset, where intent accuracy and SLURP-F1 are used as the evaluation metric. 

Our baseline model is a Conformer Transducer model, initialized from our pretrained ASR model~\footnote{\url{https://catalog.ngc.nvidia.com/orgs/nvidia/teams/nemo/models/stt_en_conformer_transducer_large}}. 
We also include results from two state-of-the-art models from ESPNet-SLU~\cite{arora2022espnet} and SpeechBrain~\cite{wang2021fine} for comparison. Both ESPNet-SLU  and SpeechBrain  use HuBERT~\cite{hsu2021hubert} encoders pretrained on LibriLight-60k~\cite{kahn2020librilight}, while ESPNet-SLU further finetunes the encoder on LibriSpeech before training on SLURP. 
For our TDTs, we use the same duration configurations that contain maximum durations 2, 4, 6, and 8. Different from our ASR and ST experiments that use $\sigma=0.05$, here we use $\sigma=0.02$ for all experiments since we found using $\sigma=0.05$ may destabilize training for SLURP\footnote{This is caused by a much smaller ratio between the audio and text length of SLU datasets: the average ratio of audio to and text is 0.89:1 for SLURP, compared to around 5.5:1 for ASR for example. Since on average audio is shorter than text, setting $\sigma$ too high, which encourages large duration outputs, will hurt training. A smaller $\sigma$  alleviates the issue.}. 

The results are shown in Table~\ref{tab:slu}. While the RNNT baseline already has better performance than ESPNet-SLU and SpeechBrain baselines,  TDT models with [0-4], [0-6], and [0-8] configurations achieve even better accuracy which makes the new state-of-the-art in the SICSF. In addition, the TDT [0-8] model is 1.28X faster in inference  than RNNT.\footnote{SLURP has on average shorter audio than text 
% with lenght ratio of 0.89:1 between audio and text
. This means larger durations occur much less for SLURP, resulting in smaller speed-ups compared to ASR and ST.}
The results demonstrate not only the effectiveness but also the efficiency of TDT algorithm applied in SICSF tasks. 

We notice relatively smaller speed-up factors with TDT in this task. This could be explained by the much lower average audio-to-token-length ratio for SICSF tasks. For example, in ASR tasks, the typical ratio between audio length to text length is around 7:1, and the ratio is around 0.89:1 for the SLURP testset, with average text sequence being longer than audio sequence\footnote{Text sequence is computed as the number of subword tokens, and audio sequence is computed as the number of 40ms frames due to 10ms per audio frame during feature extraction, with 4X subsampling performed by the encoder.}. Nevertheless, we see a significant speed up, which also shows that TDT models can bring improvement even for scenarios where the audio sequence is longer than the text sequences.

\section{Discussion}\label{analysis}

\subsection{TDT Emission Analysis}
\begin{figure}[t]
    \centering
    \includegraphics[scale=0.32]{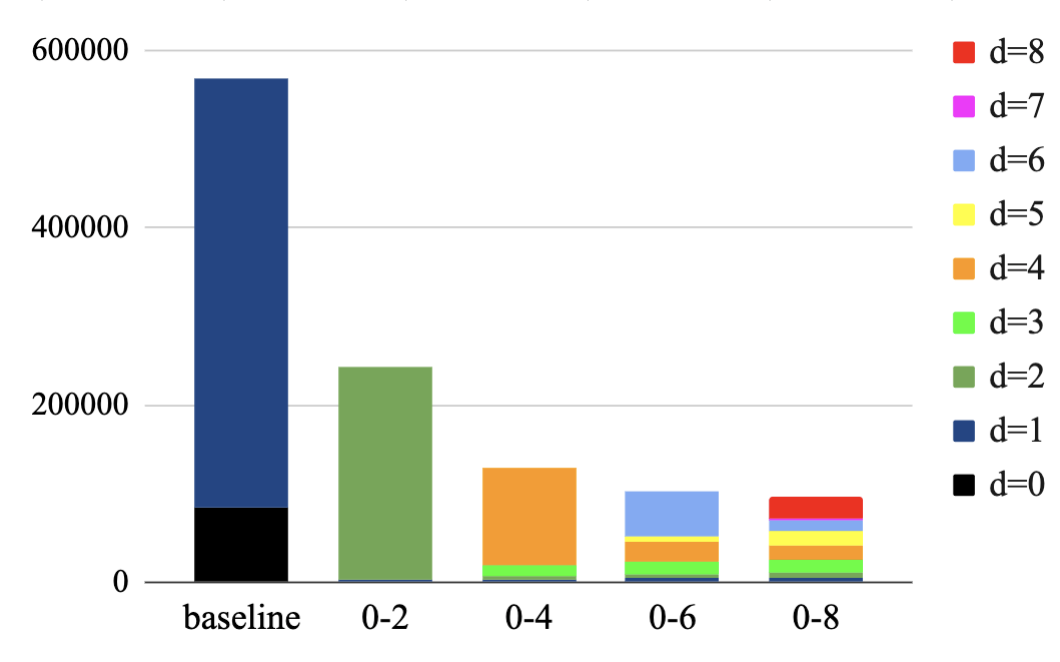}
    \caption{Duration distribution during inference on Librispeech test-other. The model is trained with 4X subsampling. The y-axis shows the number of emissions of different types of durations during the inference of Librispeech test-other datasets.}
    \label{fig:distribution}
\end{figure}

In this section, we investigate the output distribution of TDT models using Librispeech test-other dataset. 
First, we collect statistics on the duration predicted during decoding, with different TDT configurations (Fig.~\ref{fig:distribution}). For the baseline RNN-T, we treat blank and non-blank symbol emissions as having durations 1 and 0, respectively, since blank advances the $t$ by one and non-blank does not. TDT models with configs [0-2] and [0-4] fully utilize longer durations during inference, with almost all of the durations predicted for the [0-2] model, and around 90\% of durations for the [0-4] model have the maximum duration.
For [0-6] and [0-8] models, the frequencies of  predicted long durations are reduced. This is expected since our analysis shows the average ratio of audio length to text length for Librispeech test-other is 5.5:1, smaller than 6.  Hence, it is not possible to always emit such long durations.

\begin{figure}[t]
    \centering
    \includegraphics[scale=0.32]{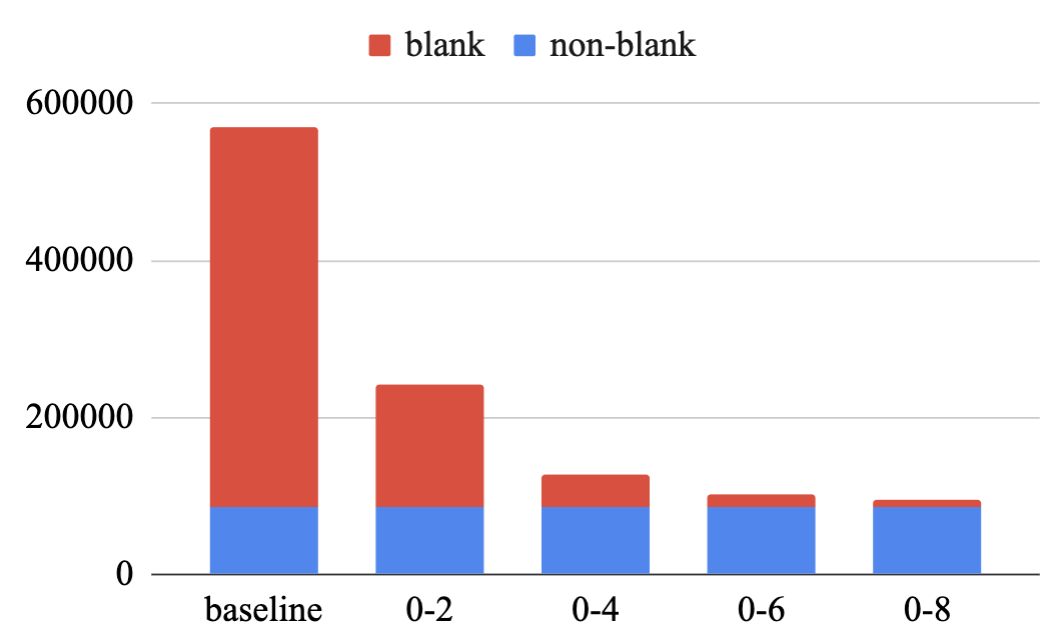}
    \caption{Emission counts of blanks VS non-blanks in Librispeech test-other. The model is trained with 4X subsampling. The y-axis shows the number of emissions of either a blank symbol (red) or a non-blank symbol (blue) during the inference of Librispeech test-other datasets.}
    \label{blank_vs_nonblank}
\end{figure}

Next, we collect the frequency of blank emissions vs non-blank emissions (Fig.~\ref{blank_vs_nonblank}). We can see that as the model incorporates longer durations, fewer and fewer blank emissions are produced with TDT models, while the number of non-blank emissions remains unchanged. TDT models with durations [0-6] and [0-8] have very few blank emissions, indicating that TDT models are close to  the theoretical lower bound in terms of the least number of decoding steps. 


%\subsection{duration distribution after blank VS non-blank emissions}

\subsection{TDT Batched Inference}\label{batch}
% \begin{table}[h!]
%     \centering
%     \begin{tabular}{c c c}
%     \toprule
%        TDT config & test-clean & test-other  \\
%     \midrule
%     baseline & 2.14 & 5.12 \\
%         0-2 & 2.34  & 5.53 \\
%         0-4 & 5.21  & 7.66 \\
%         0-6 & 7.70  & 9.62 \\
%         0-8 & 4.10  & 7.03 \\
%     \bottomrule
%     \end{tabular}
%     \caption{WER of Librispeech Test-clean and Test-other with inexact batched inference with TDT models. Batch-size is 2 for all  models. Not surprisingly, we see significant WER degradation especially with models with larger duration configurations, because such TDT models are running inference in conditions that are different from training.}
%     \label{bad_wer_sharing}
% \end{table}
% In this section, we investigate batched inference with TDT models. 
The main difficulty with batched inference for TDT models is that utterances in the batch may have different duration outputs that denote the number of frames that should be skipped. As a result, it is difficult to fully parallelize the computation for the same batch. One can make the whole batch skip the same number of frames, by selecting the minimum of predicted  durations, e.g. if batch-size=4, and predicted durations are \{3, 4, 3, 6\}, we advance the whole batch by 3 frames. However,  we found this method  results in significantly increased insertion errors with  the same tokens repeated multiple times. This is not surprising since we skip fewer frames than the model indicates, and the model would not be ready to emit the next token, but can only emit previously emitted tokens instead. 
To solve this issue, we propose a modification to the training loss of our models, by combining TDT loss with conventional transducer loss $L_\text{TDT}$ with a sampling probability $\omega$:
%\footnote{We also experimented with a linear combination of the loss function, which performs the same as sampling but is computationally more expensive to run.}, i.e.
% \begin{enumerate}
%     \item linear combination
% \begin{equation}
%     \mathcal{L} = \omega \mathcal{L_\text{transducer}} + (1 - \omega) \mathcal{L_\text{TDT}}
% \end{equation}
%     \item loss sampling
\begin{equation}
    \mathcal{L}_\text{sampled} = \begin{cases}
        \mathcal{L_\text{Transducer}}, & \text{with probability } \omega \\
        \mathcal{L_\text{TDT}}, & \text{with probability } 1 - \omega \\
    \end{cases}
\end{equation}
% \end{enumerate}
Note, the conventional transducer loss $\mathcal{L}_\text{Transducer}$ is  computed on the token logits only, and the duration logits will not take part in the computation nor get updated.
We found that the sampled loss  solves the aforementioned performance degradation issue. Table \ref{loss_combination} shows the ASR performance and inference speed of TDT models when training with $\omega = 0.1$, and running inference with batch=4. We even see slightly improved ASR accuracy, as well as inference speed-up with batched inference with TDT. \footnote{The speed-up for batched inference is slightly smaller than for non-batched case because 1. the overhead related to padding for batched computation and 2. all utterances in the batch advance by the minimum of predicted durations which increases the number of decoding steps.}
\begin{table}[h!]
    \centering
    \begin{tabular}{c c c c c}
    \toprule
       TDT config  & clean & other  & total time & rel speed-up\\
       \midrule
        RNNT & 2.13 & 5.11 & 274 & -     \\
        
        0-2     & 2.10 & 4.94 & 182 & 1.51X \\
        0-4     & 2.15 & 5.04 & 151 & 1.81X \\
        0-6     & 2.10 & 4.91 & 146 & 1.88X \\
        0-8     & 2.13 & 5.03 & 159 & 1.79X \\   
    \bottomrule
    \end{tabular}
    \caption{Batched inference for TDT ASR models, trained with loss sampling $\omega = 0.1$. WER (\%) on Librispeech test-clean and test-other. Batch-size=4. When different utterances in the same batch predict different durations, we take the minimum of those predictions and advance all utterances in the batch by that amount.
    }
    \label{loss_combination}
\end{table}

\subsection{TDT Robustness to noise}
\begin{figure}[h!]
    \centering
    \includegraphics[scale=0.32]{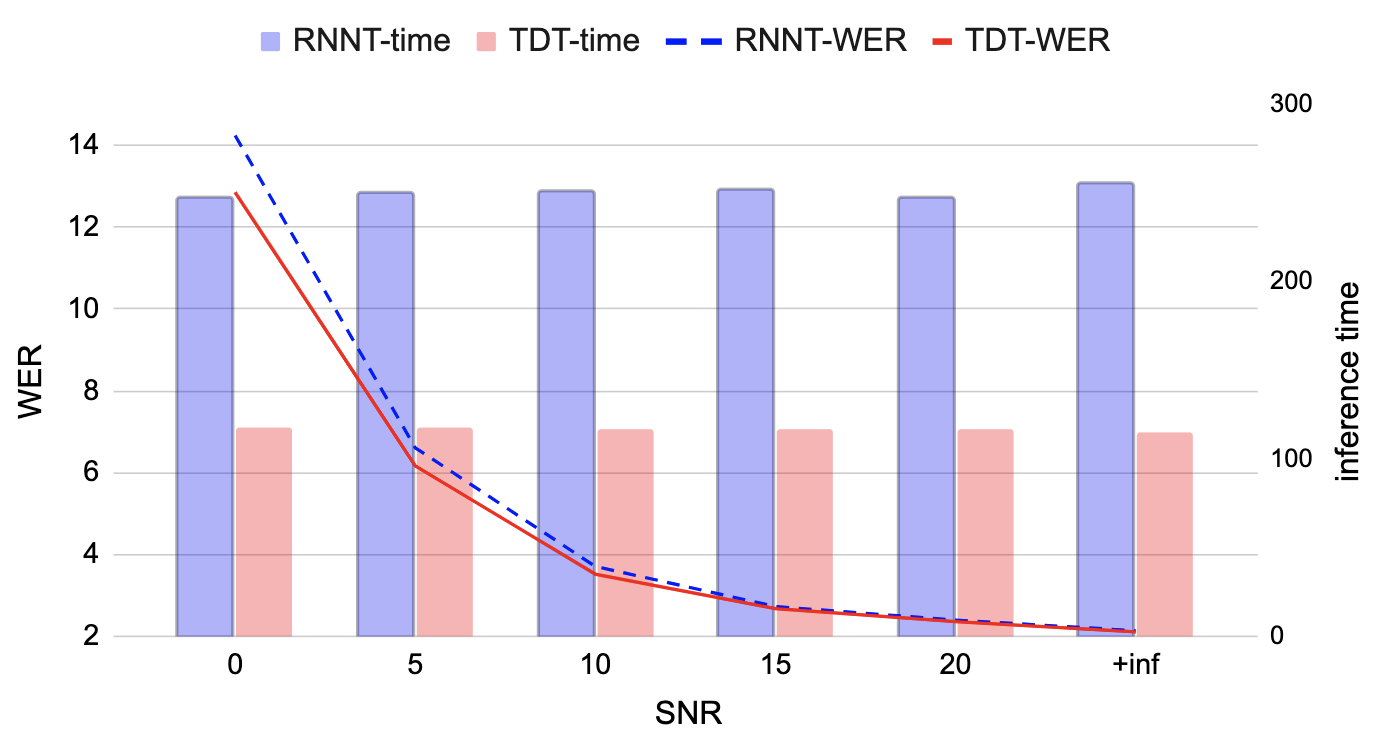}
    \caption{TDT vs RNNT ASR on noisy speech. WER(\%) for Librispeech test-clean with noise added at different SNRs. WER on the original test-clean is shown at SNR = +inf. While TDT and RNNT achieve similar WER at low noise conditions, TDT is more robust to noise.}
    \label{SNR}
\end{figure}

In this section, we compare the noise robustness for TDT and RNNT ASR models. For this, we run inference on Librispeech test-clean augmented with noise in different signal-noise-ratios (SNRs). 
For each utterance, we randomly select a noise sample from MUSAN~\cite{snyder2015musan} and Freesound~\footnote{\url{https://freesound.org/}}. 
The noise sample is sub-segmented if it's longer than the utterance, or repeated if it's shorter than the utterance. The utterance samples are augmented with noise samples in 0, 5, 10, 15, and 20 SNRs.
We report the WER and inference time of conventional Transducers and TDT models with configuration 0-8.
We found TDT models perform much better in noisy conditions than conventional Transducers, both in terms of accuracy and speed (Fig.~\ref{SNR}). While RNN-T and TDT models achieve similar WERs for clean speech, TDT models gradually outperform RNNT as more noise is added.  The inference time for TDT is practically the same for all SNRs. More details of those experiments are in Appendix E.

\subsection{TDT Robustness with respect to repeated tokens}
We notice that RNN-T model performance significantly degrades when the text sequence has repetitions of the same (subword) tokens, for example:
\begin{itemize}
    \item Ground truth: \emph{seven seven seven nine nine nine eight eight eight}
    \item RNNT w/ LSTM decoder result: \emph{seven seven eight eight}
    \item RNNT w/ stateless decoder result: \emph{seven nine eight}
    % \item RNNT-TDT 0-8: seven seven seven nine nine nine eight eight eight
\end{itemize}
We find TDT models are significantly more robust than RNN-Ts for such cases. 
We use NeMo TTS to generate 100 audios containing random digits repeating 3 - 5 times and run ASR with different models with results in Table \ref{repeated_wer}. We see that while the conventional RNN-Ts achieve very bad word error rates (all of them with error rates more than 50\%, regardless of the type of decoder used), TDT models are able to achieve significantly lower error rates, and this effect is more prominent for TDT models with longer durations. With TDT models with durations 0-8, we achieve a word error rate of 5.78\%, which is more than 10X error rate reduction compared to conventional Transducers. 

This set of experiments also shows that TDT models do not suffer from the potential issue of accumulation of errors induced by consecutive duration predictions that might be inaccurate.
More details on the analysis of repeated tokens and our experiments can be found in Appendix F. 

\begin{table}[]
    \centering
    \begin{tabular}{c c}
    \toprule
        model & WER\%\\
    \midrule
        RNNT-LSTM & 59.95 \\
        RNNT-stateless & 64.62 \\
        TDT [0-2] & 12.59 \\
        TDT [0-4] & 9.35 \\
        TDT [0-6] & 6.12 \\
        TDT [0-8] & 5.78 \\
    \bottomrule
    \end{tabular}
    \caption{WERs with different Transducer models on TTS generated dataset with repeated digits.}
    \label{repeated_wer}
\end{table}

\subsection{TDT Comparison with Multi-blank Transducers}
Multi-blank Transducer (MBT) \cite{xu2022multi} introduces \emph{big blank} symbols that cover multiple input frames. During inference, when a big blank is emitted, the MBT model skips frames according to the duration of the emitted blank symbol. Compared to multi-blank Transducers which skip frames only with certain blank symbols,  TDT models allow frame-skipping for both non-blank and blank symbols,  which means potentially larger speed-up factors. We compare MBT and TDT in Table \ref{compare_multiblank}. We see that while both MBT and TDT models give comparable WERs, larger inference speedup factors  are seen with TDT models when using the same max-duration configs. %, and thus TDT models are superior to MBT models.
\begin{table}[t]
    \centering
    \begin{tabular}{c c c c c}
    \toprule
    model & max-duration & WER & time & rel. speed up \\
    \midrule
    RNNT        & - & 5.11 & 244 & - \\
    \midrule
    MBT              & 2     & 5.15  & 208 &  1.17X \\
    TDT              & 2     & 5.50 & 171 & 1.43X \\   
    \midrule
    MBT              & 4     & 5.05 & 161 & 1.52X \\
    TDT              & 4     & 5.06 & 128 & 1.91X  \\
    \midrule
    MBT              & 8     & 5.18 & 139 & 1.76X \\
    TDT              & 8     & 5.16 & 115 & 2.12X \\
    \bottomrule
    \end{tabular}
    \caption{Inference and accuracy comparison between 3 type of ASR models: RNNT, multi-blank RNNT (MBT), and TDT. Greedy WER (\%) and the total decoding time of the Librispeech test-other with batch = 1. Relative speed-up is measured against the RNNT. For MBT max-duration=4 means MBT model with big-blank-durations=[2,3,4] in addition to the conventional blank. For TDT models,  max-duration=4 means model with durations [0,1,2,3,4].}
    \label{compare_multiblank}
\end{table}

\section{Conclusion}\label{future}
In this paper, we propose Token-and-Duration Transducers, which extend conventional Transducer models by adding explicit duration modeling.
We present detailed derivations of the extended forward-backward algorithm used for TDT models, as well as the close-form solutions for TDT model training. %, including
%gradients of pre-softmax logits for token outputs inspired by function-merging \cite{li2019improving} methods. 
We show that TDT models are superior to conventional Transducers across multiple sequence tasks, including speech recognition, speech translation, and spoken language understanding. In all those tasks, we see better or similar performances with TDT models than conventional Transducers, while TDT models run inference significantly faster, with up to 2.82X speed up. TDT is also more noise-robust, and robust to token repetition than conventional RNN-Ts.
Our TDT implementation  will be open-sourced in NVIDIA's NeMo \url{https://github.com/NVIDIA/NeMo} toolkit.

For future work, we will work on other ways to improve the computational efficiency and accuracy of TDT models, as well as algorithms and implementation for efficient beam-search with TDT models. 

% With speech recognition experiments on four languages and multiple datasets, we show that TDT models achieve slight gains in terms of accuracy, and bring massive speed-up in model inference, as much as 2.82X in the case of German Multilingual Librispeech when running non-batched inference. The proposed model is ideal for use in on-device speech recognition systems for its vast performance gain in non-batched inference. We also propose a special training method, which allows TDT models to run efficient batch inference that can be run on servers.

% For future work, we plan to further improve TDT models and apply similar methods to other types of models, e.g. CTC models.

\nocite{langley00}

\bibliography{main}

\begin{thebibliography}{41}
\providecommand{\natexlab}[1]{#1}
\providecommand{\url}[1]{\texttt{#1}}
\expandafter\ifx\csname urlstyle\endcsname\relax
  \providecommand{\doi}[1]{doi: #1}\else
  \providecommand{\doi}{doi: \begingroup \urlstyle{rm}\Url}\fi

\bibitem[Ardila et~al.(2019)Ardila, Branson, Davis, Henretty, Kohler, Meyer,
  Morais, Saunders, Tyers, and Weber]{ardila2019common}
Ardila, R., Branson, M., Davis, K., Henretty, M., Kohler, M., Meyer, J.,
  Morais, R., Saunders, L., Tyers, F.~M., and Weber, G.
\newblock Common voice: A massively-multilingual speech corpus.
\newblock \emph{arXiv:1912.06670}, 2019.

\bibitem[Arora et~al.(2022)Arora, Dalmia, Denisov, Chang, Ueda, Peng, Zhang,
  Kumar, Ganesan, Yan, et~al.]{arora2022espnet}
Arora, S., Dalmia, S., Denisov, P., Chang, X., Ueda, Y., Peng, Y., Zhang, Y.,
  Kumar, S., Ganesan, K., Yan, B., et~al.
\newblock {ESPnet-SLU}: Advancing spoken language understanding through
  {ESPnet}.
\newblock In \emph{ICASSP}, 2022.

\bibitem[Bastianelli et~al.(2020)Bastianelli, Vanzo, Swietojanski, and
  Rieser]{slurp}
Bastianelli, E., Vanzo, A., Swietojanski, P., and Rieser, V.
\newblock {SLURP: A Spoken Language Understanding Resource Package}.
\newblock In \emph{EMNLP}, 2020.

\bibitem[Cattoni et~al.(2021)Cattoni, Di~Gangi, Bentivogli, Negri, and
  Turchi]{cattoni2021must}
Cattoni, R., Di~Gangi, M.~A., Bentivogli, L., Negri, M., and Turchi, M.
\newblock Must-c: A multilingual corpus for end-to-end speech translation.
\newblock \emph{Computer Speech \& Language}, 66:\penalty0 101155, 2021.

\bibitem[Chan et~al.(2016)Chan, Jaitly, Le, and Vinyals]{chan2015listen}
Chan, W., Jaitly, N., Le, Q.~V., and Vinyals, O.
\newblock {Listen, Attend and Spell}.
\newblock In \emph{ICASSP}, 2016.

\bibitem[Chorowski et~al.(2015)Chorowski, Bahdanau, Serdyuk, Cho, and
  Bengio]{chorowski2015attention}
Chorowski, J.~K., Bahdanau, D., Serdyuk, D., Cho, K., and Bengio, Y.
\newblock Attention-based models for speech recognition.
\newblock \emph{NeurIPS}, 28, 2015.

\bibitem[Farhad et~al.(2021)Farhad, Arkady, Magdalena, Ond{\v{r}}ej, Rajen,
  Vishrav, Costa-jussa, Cristina, Angela, Christian,
  et~al.]{farhad2021findings}
Farhad, A., Arkady, A., Magdalena, B., Ond{\v{r}}ej, B., Rajen, C., Vishrav,
  C., Costa-jussa, M.~R., Cristina, E.-B., Angela, F., Christian, F., et~al.
\newblock Findings of the 2021 conference on machine translation ({WMT21}).
\newblock In \emph{Conf. on Machine Translation}, 2021.

\bibitem[Ghodsi et~al.(2020)Ghodsi, Liu, Apfel, Cabrera, and
  Weinstein]{Ghodsi2020stateless}
Ghodsi, M., Liu, X., Apfel, J., Cabrera, R., and Weinstein, E.
\newblock {RNN-Transducer} with stateless prediction network.
\newblock In \emph{ICASSP}, 2020.

\bibitem[Graves(2012)]{graves2012sequence}
Graves, A.
\newblock Sequence transduction with recurrent neural networks.
\newblock In \emph{ICML}, 2012.

\bibitem[Graves et~al.(2006)Graves, Fern{\'a}ndez, Gomez, and
  Schmidhuber]{graves2006connectionist}
Graves, A., Fern{\'a}ndez, S., Gomez, F., and Schmidhuber, J.
\newblock Connectionist temporal classification: labelling unsegmented sequence
  data with recurrent neural networks.
\newblock In \emph{ICML}, 2006.

\bibitem[Gulati et~al.(2020)Gulati, Qin, Chiu, Parmar, Zhang, Yu, Han, Wang,
  Zhang, Wu, et~al.]{gulati2020conformer}
Gulati, A., Qin, J., Chiu, C.-C., Parmar, N., Zhang, Y., Yu, J., Han, W., Wang,
  S., Zhang, Z., Wu, Y., et~al.
\newblock Conformer: Convolution-augmented transformer for speech recognition.
\newblock In \emph{Interspeech}, 2020.

\bibitem[Han et~al.(2020)Han, Zhang, Zhang, Yu, Chiu, Qin, Gulati, Pang, and
  Wu]{contextnet}
Han, W., Zhang, Z., Zhang, Y., Yu, J., Chiu, C.-C., Qin, J., Gulati, A., Pang,
  R., and Wu, Y.
\newblock Contextnet: Improving convolutional neural networks for automatic
  speech recognition with global context.
\newblock In \emph{Interspeech}, 2020.

\bibitem[Hsu et~al.(2021)Hsu, Bolte, Tsai, Lakhotia, Salakhutdinov, and
  Mohamed]{hsu2021hubert}
Hsu, W.-N., Bolte, B., Tsai, Y.-H.~H., Lakhotia, K., Salakhutdinov, R., and
  Mohamed, A.
\newblock Hubert: Self-supervised speech representation learning by masked
  prediction of hidden units.
\newblock \emph{IEEE/ACM Transactions on Audio, Speech, and Language
  Processing}, 29:\penalty0 3451--3460, 2021.

\bibitem[Indurthi et~al.(2021)Indurthi, Zaidi, Lakumarapu, Lee, Han, Ahn, Kim,
  Kim, and Hwang]{indurthi2021task}
Indurthi, S., Zaidi, M.~A., Lakumarapu, N.~K., Lee, B., Han, H., Ahn, S., Kim,
  S., Kim, C., and Hwang, I.
\newblock Task aware multi-task learning for speech to text tasks.
\newblock In \emph{ICASSP}, 2021.

\bibitem[Iranzo-S{\'a}nchez et~al.(2020)Iranzo-S{\'a}nchez, Silvestre-Cerda,
  Jorge, Rosell{\'o}, Gim{\'e}nez, Sanchis, Civera, and
  Juan]{iranzo2020europarl}
Iranzo-S{\'a}nchez, J., Silvestre-Cerda, J.~A., Jorge, J., Rosell{\'o}, N.,
  Gim{\'e}nez, A., Sanchis, A., Civera, J., and Juan, A.
\newblock Europarl-st: A multilingual corpus for speech translation of
  parliamentary debates.
\newblock In \emph{ICASSP}, 2020.

\bibitem[Jelinek(1998)]{jelinek1998statistical}
Jelinek, F.
\newblock \emph{Statistical methods for speech recognition}.
\newblock MIT press, 1998.

\bibitem[Kahn et~al.(2020)Kahn, Rivi{\`e}re, Zheng, Kharitonov, Xu, Mazar{\'e},
  Karadayi, Liptchinsky, Collobert, Fuegen, et~al.]{kahn2020librilight}
Kahn, J., Rivi{\`e}re, M., Zheng, W., Kharitonov, E., Xu, Q., Mazar{\'e},
  P.-E., Karadayi, J., Liptchinsky, V., Collobert, R., Fuegen, C., et~al.
\newblock Libri-light: A benchmark for {ASR} with limited or no supervision.
\newblock In \emph{ICASSP}, 2020.

\bibitem[Kingma \& Ba(2015)Kingma and Ba]{kingma_adam}
Kingma, D.~P. and Ba, J.
\newblock Adam: A method for stochastic optimization.
\newblock In \emph{ICLR}, 2015.

\bibitem[Kuchaiev et~al.(2019)Kuchaiev, Li, Nguyen, Hrinchuk, Leary, Ginsburg,
  Kriman, Beliaev, Lavrukhin, et~al.]{kuchaiev2019nemo}
Kuchaiev, O., Li, J., Nguyen, H., Hrinchuk, O., Leary, R., Ginsburg, B.,
  Kriman, S., Beliaev, S., Lavrukhin, V., et~al.
\newblock Nemo: a toolkit for building ai applications using neural modules.
\newblock In \emph{{NeurIPS Workshop on Systems for ML}}, 2019.

\bibitem[Li et~al.(2019)Li, Zhao, Hu, and Gong]{li2019improving}
Li, J., Zhao, R., Hu, H., and Gong, Y.
\newblock Improving {RNN} transducer modeling for end-to-end speech
  recognition.
\newblock In \emph{ASRU}, 2019.

\bibitem[Niehues et~al.(2018)Niehues, Cattoni, St{\"u}ker, Cettolo, Turchi, and
  Federico]{niehues-etal-2018-iwslt}
Niehues, J., Cattoni, R., St{\"u}ker, S., Cettolo, M., Turchi, M., and
  Federico, M.
\newblock The {IWSLT} 2018 evaluation campaign.
\newblock In \emph{IWSLT}, 2018.

\bibitem[Panayotov et~al.(2015)Panayotov, Chen, Povey, and
  Khudanpur]{panayotov2015librispeech}
Panayotov, V., Chen, G., Povey, D., and Khudanpur, S.
\newblock Librispeech: an {ASR} corpus based on public domain audio books.
\newblock In \emph{ICASSP}, 2015.

\bibitem[Paszke et~al.(2019)]{paszke_pytorch}
Paszke, A. et~al.
\newblock {PyTorch: An Imperative Style, High-Performance Deep Learning
  Library}.
\newblock In \emph{NeuRIPS}, 2019.

\bibitem[Povey et~al.(2011)Povey, Ghoshal, Boulianne, Burget, Glembek, Goel,
  Hannemann, Motlicek, Qian, Schwarz, Silovsky, Stemmer, and
  Vesely]{povey2011kaldi}
Povey, D., Ghoshal, A., Boulianne, G., Burget, L., Glembek, O., Goel, N.,
  Hannemann, M., Motlicek, P., Qian, Y., Schwarz, P., Silovsky, J., Stemmer,
  G., and Vesely, K.
\newblock The {Kaldi} speech recognition toolkit.
\newblock In \emph{ASRU}, 2011.

\bibitem[Pratap et~al.(2020)Pratap, Xu, Sriram, Synnaeve, and
  Collobert]{pratap2020mls}
Pratap, V., Xu, Q., Sriram, A., Synnaeve, G., and Collobert, R.
\newblock {MLS}: A large-scale multilingual dataset for speech research.
\newblock In \emph{Interspeech}, 2020.

\bibitem[Ravanelli et~al.(2021)Ravanelli, Parcollet, Plantinga, Rouhe, Cornell,
  Lugosch, Subakan, Dawalatabad, Heba, Zhong, et~al.]{ravanelli2021speechbrain}
Ravanelli, M., Parcollet, T., Plantinga, P., Rouhe, A., Cornell, S., Lugosch,
  L., Subakan, C., Dawalatabad, N., Heba, A., Zhong, J., et~al.
\newblock {SpeechBrain}: A general-purpose speech toolkit.
\newblock In \emph{Interspeech}, 2021.

\bibitem[Sennrich et~al.(2015)Sennrich, Haddow, and Birch]{sennrich2015neural}
Sennrich, R., Haddow, B., and Birch, A.
\newblock Neural machine translation of rare words with subword units.
\newblock In \emph{Proc. of the 54th Annual Meeting of the ACL}, 2015.

\bibitem[Shrivastava et~al.(2021)Shrivastava, Garg, Cao, Zhang, and
  Sainath]{shrivastava2021echo}
Shrivastava, H., Garg, A., Cao, Y., Zhang, Y., and Sainath, T.
\newblock Echo state speech recognition.
\newblock In \emph{ICASSP}, 2021.

\bibitem[Snyder et~al.(2015)Snyder, Chen, and Povey]{snyder2015musan}
Snyder, D., Chen, G., and Povey, D.
\newblock Musan: A music, speech, and noise corpus.
\newblock \emph{arXiv:1510.08484}, 2015.

\bibitem[Tian et~al.(2019)Tian, Yi, Tao, Bai, and Wen]{tian2019self}
Tian, Z., Yi, J., Tao, J., Bai, Y., and Wen, Z.
\newblock Self-attention transducers for end-to-end speech recognition.
\newblock In \emph{Interspeech}, 2019.

\bibitem[Wang et~al.(2021{\natexlab{a}})Wang, Riviere, Lee, Wu, Talnikar,
  Haziza, Williamson, Pino, and Dupoux]{wang2021voxpopuli}
Wang, C., Riviere, M., Lee, A., Wu, A., Talnikar, C., Haziza, D., Williamson,
  M., Pino, J., and Dupoux, E.
\newblock {VoxPopuli}: A large-scale multilingual speech corpus for
  representation learning, semi-supervised learning and interpretation.
\newblock In \emph{Proc. of the 59th Annual Meeting of the ACL and the 11th
  International Joint Conf. on NLP}, 2021{\natexlab{a}}.

\bibitem[Wang et~al.(2021{\natexlab{b}})Wang, Wu, and Pino]{wang2020covost}
Wang, C., Wu, A., and Pino, J.
\newblock {CoVoST} 2 and massively multilingual speech-to-text translation.
\newblock In \emph{Interspeech}, 2021{\natexlab{b}}.

\bibitem[Wang et~al.(2019)Wang, Chen, Xu, Ding, Lv, Shao, Peng, Xie, Watanabe,
  and Khudanpur]{wang2019espresso}
Wang, Y., Chen, T., Xu, H., Ding, S., Lv, H., Shao, Y., Peng, N., Xie, L.,
  Watanabe, S., and Khudanpur, S.
\newblock Espresso: A fast end-to-end neural speech recognition toolkit.
\newblock In \emph{ASRU}, 2019.

\bibitem[Wang et~al.(2021{\natexlab{c}})Wang, Boumadane, and
  Heba]{wang2021fine}
Wang, Y., Boumadane, A., and Heba, A.
\newblock A fine-tuned {Wav2vec 2.0/HuBERT} benchmark for speech emotion
  recognition, speaker verification and spoken language understanding.
\newblock \emph{arXiv:2111.02735}, 2021{\natexlab{c}}.

\bibitem[Watanabe et~al.(2018)Watanabe, Hori, Karita, Hayashi, Nishitoba, Unno,
  {Enrique Yalta Soplin}, Heymann, Wiesner, Chen, Renduchintala, and
  Ochiai]{watanabe2018espnet}
Watanabe, S., Hori, T., Karita, S., Hayashi, T., Nishitoba, J., Unno, Y.,
  {Enrique Yalta Soplin}, N., Heymann, J., Wiesner, M., Chen, N.,
  Renduchintala, A., and Ochiai, T.
\newblock {ESPnet}: End-to-end speech processing toolkit.
\newblock In \emph{Interspeech}, 2018.

\bibitem[Woodland et~al.(1994)Woodland, Odell, Valtchev, and
  Young]{woodland1994large}
Woodland, P.~C., Odell, J.~J., Valtchev, V., and Young, S.~J.
\newblock Large vocabulary continuous speech recognition using {HTK}.
\newblock In \emph{ICASSP}, 1994.

\bibitem[Xu et~al.(2022)Xu, Jia, Majumdar, Watanabe, and Ginsburg]{xu2022multi}
Xu, H., Jia, F., Majumdar, S., Watanabe, S., and Ginsburg, B.
\newblock Multi-blank transducers for speech recognition.
\newblock \emph{arXiv:2211.03541}, 2022.

\bibitem[Xue et~al.(2022)Xue, Wang, Li, Post, and Gaur]{xue2022large}
Xue, J., Wang, P., Li, J., Post, M., and Gaur, Y.
\newblock Large-scale streaming end-to-end speech translation with neural
  transducers.
\newblock \emph{arXiv preprint arXiv:2204.05352}, 2022.

\bibitem[Yeh et~al.(2019)Yeh, Mahadeokar, Kalgaonkar, Wang, Le, Jain, Schubert,
  Fuegen, and Seltzer]{yeh2019transformer}
Yeh, C.-F., Mahadeokar, J., Kalgaonkar, K., Wang, Y., Le, D., Jain, M.,
  Schubert, K., Fuegen, C., and Seltzer, M.~L.
\newblock Transformer-transducer: End-to-end speech recognition with
  self-attention.
\newblock \emph{arXiv:1910.12977}, 2019.

\bibitem[Yu et~al.(2021)Yu, Chiu, Li, Chang, Sainath, He, Narayanan, Han,
  Gulati, Wu, et~al.]{yu2021fastemit}
Yu, J., Chiu, C.-C., Li, B., Chang, S.-y., Sainath, T.~N., He, Y., Narayanan,
  A., Han, W., Gulati, A., Wu, Y., et~al.
\newblock {FastEmit}: Low-latency streaming {ASR} with sequence-level emission
  regularization.
\newblock In \emph{ICASSP}, 2021.

\bibitem[Zhang et~al.(2020)Zhang, Lu, Sak, Tripathi, McDermott, Koo, and
  Kumar]{zhang2020transformer}
Zhang, Q., Lu, H., Sak, H., Tripathi, A., McDermott, E., Koo, S., and Kumar, S.
\newblock {Transformer Transducer}: A streamable speech recognition model with
  transformer encoders and {RNN-T} loss.
\newblock In \emph{ICASSP}, 2020.

\end{thebibliography}
\bibliographystyle{icml2021}

%%%%%%%%%%%%%%%%%%%%%%%%%%%%%%%%%%%%%%%%%%%%%%%%%%%%%%%%%%%%%%%%%%%%%%%%%%%%%%%
%%%%%%%%%%%%%%%%%%%%%%%%%%%%%%%%%%%%%%%%%%%%%%%%%%%%%%%%%%%%%%%%%%%%%%%%%%%%%%%
% APPENDIX
%%%%%%%%%%%%%%%%%%%%%%%%%%%%%%%%%%%%%%%%%%%%%%%%%%%%%%%%%%%%%%%%%%%%%%%%%%%%%%%
%%%%%%%%%%%%%%%%%%%%%%%%%%%%%%%%%%%%%%%%%%%%%%%%%%%%%%%%%%%%%%%%%%%%%%%%%%%%%%%
\newpage
\appendix
\onecolumn

\section{ Derivations of TDT gradients with respect to probabilities}
\subsection{Background: gradients of RNN-Transducer}
The gradient of the conventional Transducer loss 
function with respect to $P(v|t, u)$ has been derived in \cite{graves2012sequence}.  The total probability $p(y | x)$ equals to the sum of $\alpha(t, u) \beta(t, u)$ over any top-left to bottom-right diagonal through the probability lattice, i.e. $\forall n: 1 \leq n \leq U + T$
\begin{equation} 
\label{diagonal}
    P_\text{RNNT}(y|x) = \sum_{(t,u): t + u =n} \alpha(t, u) \beta(t, u)
\end{equation}
We plug $\beta(t, u) = \beta(t + 1, u) p(\O | t, u) + \beta(t, u + 1) p(y_u | t, u)$, in the equation above and get: 
\begin{equation}
    P_\text{RNNT}(y|x) = \sum_{(t,u): t + u =n} \alpha(t, u) \Biggl[  \beta(t + 1, u) p(\O | t, u) + \beta(t, u + 1) p(y_{u+1} | t, u) \Biggr]
\end{equation}
Now let us take the partial derivative of $p(y|x)$ over individual $p(v | t, u)$: 
\begin{equation}
    \frac{\partial P_\text{RNNT}(y | x)}{\partial P(v | t, u)} =  \alpha(t, u) \begin{cases}
        \beta(t,u+1),  & v = y_{u+1} \\
        \beta(t+1,u),  & v = \O \\
        0,             & \text{otherwise.} \\
    \end{cases}
\end{equation}
Since $\mathcal{L}_\text{RNNT} =  -\log P_\text{RNNT}(y | x)$, we have
\begin{equation}
\label{rnnt_grad}
    \frac{\partial \mathcal{L}_\text{RNNT}}{\partial P(v|t, u)} = \frac{d \mathcal{L}_\text{RNNT}}{d P_\text{RNNT}(y | x)} \frac{\partial {P_\text{RNNT}(y | x)}}{{\partial P(v|t, u)}} =   -\frac{\alpha(t,u)}{P_\text{RNNT}(y|x)}\begin{cases}
        \beta(t,u+1),  & v = y_{u+1} \\
        \beta(t+1,u),  & v = \O \\
        0,             & \text{otherwise.} \\
    \end{cases}
\end{equation}

\subsection{Gradients of TDT models}

Derivation of gradients for TDT follows similar steps as gradients for conventional Transducers. But first note, that Eq.~\ref{diagonal} does not hold true for TDT: since frame-skipping is not possible for a RNNT, for any $n$, the diagonal $(t, u): t + u$ contains a set of nodes that \emph{all} possible paths in the lattice must go through. But for TDT, frame-skipping is allowed, so we also need to consider paths that \emph{jump over} the diagonal. The correct formulation for TDT is that $\forall n: 1 \leq n \leq U + T$: 
\begin{equation}
\label{diagonal2}
\begin{aligned}
        P_\text{TDT}(y|x) & = \sum_{(t,u): t + u =n} \alpha(t, u) \beta(t, u) 
         + \sum_{(t,u,t'): t+u < n, t'+u > n, (t'-t) \in \mathcal{D}} \alpha(t, u) \beta(t', u) P(\O, t'-t | t, u) \\
        & + \sum_{(t,u,t'): t+u < n, t'+u+1 > n, (t'-t) \in \mathcal{D}} \alpha(t, u) \beta(t', u+1) P(y_{u+1}, t'-t | t, u)
\end{aligned}
\end{equation}
Eq.~\ref{diagonal2} has extra terms comparing to Eq.~\ref{diagonal}, which correspond to transitions that go across the specified diagonal.
Starting from Eq.~\ref{diagonal2}, we follow similar steps as conventional Transducers.
To get the gradients of the token probabilities, we replace $\beta(t, u)$ according to 
Eq.~\ref{eqn:betas},  and take the partial derivative of token probabilities. The last two terms do not have contributions to the partial derivative, and we obtain:
\begin{equation} 
    \label{tdt_p_gradient}
    \frac{\partial P_\text{TDT}(y|x)}{\partial P_T(v|t, u)} = \alpha(t,u)  \begin{cases}
        \mathlarger\sum\limits_{d\in \mathcal{D}} \beta(t+d,u+1) P_D(d | t, u),  & v = y_{u+1} \\
        \mathlarger\sum\limits_{d\in \mathcal{D} \setminus{\{0\} }} \beta(t+d,u) P_D(d| t, u),  & v = \O \\
        0,             & \text{otherwise.} \\
    \end{cases}
\end{equation}
Then taking $\mathcal{L}_\text{TDT} =  -\log P_\text{TDT}(y | x)$, we have:
\begin{equation} 
    \label{tdt_gradient_token_all}
    \frac{\partial \mathcal{L}_\text{TDT}}{\partial P_T(v|t, u)} = -\frac{\alpha(t,u) }{P_\text{TDT}(y|x)} \begin{cases}
        \mathlarger\sum\limits_{d\in \mathcal{D}} \beta(t+d,u+1) P_D(d | t, u),  & v = y_{u+1} \\
        \mathlarger\sum\limits_{d\in \mathcal{D} \setminus{\{0\} }} \beta(t+d,u) P_D(d| t, u),  & v = \O \\
        0,             & \text{otherwise.} \\
    \end{cases}
\end{equation}
Similarly, starting with Eq.~\ref{diagonal2} and Eq.~\ref{eqn:betas}, we take the partial derivatives of the duration probabilities:
\begin{equation}
    \label{duration_p}
    \frac{\partial P_\text{TDT}(y | x)} {\partial P_D(d|t,u)} = \alpha(t, u)  \begin{cases}
        \beta(t, u + 1) P_T(y_{u+1} | t, u), & d = 0 \\
        \beta(t + d, u + 1) P_T(y_{u+1} | t, u)  + \beta(t + d, u) P_T(\O | t, u), & d > 0. \\
    \end{cases}
\end{equation}
Since $\mathcal{L}_\text{TDT} =  -\log P_\text{TDT}(y | x)$, we get:
\begin{equation}
    \label{duration_all}
    \frac{\partial \mathcal{L}_\text{TDT}} {\partial P_D(d|t,u)} = -\frac{\alpha(t, u) }{P_\text{TDT}(y|x)} \begin{cases}
        \beta(t, u + 1) P_T(y_{u+1} | t, u), & d = 0 \\
        \beta(t + d, u + 1) P_T(y_{u+1} | t, u)  + \beta(t + d, u) P_T(\O | t, u), & d > 0. \\
    \end{cases}
\end{equation}

% Similar to \cite{graves2012sequence}, given an input sequence $\boldsymbol{x}$ and a target sequence $\boldsymbol{y^*}$, the model can be trained by minimizing the log-loss $\mathcal{L} = \text{-ln P} \left( \boldsymbol{y^*} | \boldsymbol{x} \right)$ of the target sequence. $P(\boldsymbol{y^*} | \boldsymbol{x} )$ is computed as the sum of $\alpha(t, u) \beta(t, u)$ over all top-left to bottom-right diagonal through the nodes similar to the conventional Transducer loss. At any given node, let $\forall \: n: 1 \le n \le T + U$, then 

% % \begin{equation} 
% %     \label{eqn:prob_token}
% %     P_T(y^* | x) = \mathlarger\sum_{(t, u):t+u=n} \mathlarger\sum_{v} \alpha(t, u) B(v, t, u)
% % \end{equation}
% % \begin{equation}
% %     \label{eqn:prob_duration}
% %     P_D(y^* | x) = \mathlarger\sum_{(t, u):t+u=n} \mathlarger\sum_{v} \alpha(t, u) C(v, t, u)
% % \end{equation}
% which defines the probability of all complete alignments $P(\mathcal{A}_{t, u} | x)$ at each node $(t, u)$.
% %

\section{ Derivations of TDT gradients with respect to pre-softmax logits}

\subsection{Background: Softmax function merging for conventional Transducers}
As noted previously, $h_{t, u}^v$ are the pre-softmax logits joint network for the token prediction. 
\cite{li2019improving} proposed to compute a closed-form solution of Transducer loss of $h_{t, u}^v$ by the following steps. 

First, we apply the chain rule
to represent the gradients to pre-softmax logits as follows, \footnote{In the original paper \cite{li2019improving}, although their final result is correct, there seems to be a small issue in their Eq. 12, where the authors did not include the summation.}
\begin{align}
\label{function_merging_all}
    \frac{\partial{\mathcal{L}_\text{RNNT}}}{\partial{h_{t, u}^v}} &= \sum_{v' \in \mathcal{V}  } \frac{\partial{\mathcal{L}_\text{RNNT}}}{\partial{P(v'| t, u)}} \frac{\partial{P(v'| t, u)}}{\partial{h_{t, u}^v}} %+ \frac{\partial{L}}{\partial{P_D(d| t, u)}} \frac{\partial{P_D(d| t, u)}}{\partial{h_{t, u}^v}}
\end{align}
where $\mathcal{V}$ represents a set of all vocabulary including the $\O$ token.
In the summation, $\frac{\partial{\mathcal{L}}}{\partial{P(v| t, u)}}$ can be calculated via Eq.~\ref{rnnt_grad}. Although this summation is over all elements in $\mathcal{V}$, only two of them are non-zero: the one where $v' = \O$ and where $v' = y_{u+1}$. Therefore, we could simplify Eq.~\ref{function_merging_all} as, 
\begin{equation}
\label{2terms}
\frac{\partial{\mathcal{L}_\text{RNNT}}}{\partial{h_{t, u}^v}} = \frac{\partial{\mathcal{L}_\text{RNNT}}}{\partial{P(\O| t, u)}} \frac{\partial{P(\O| t, u)}}{\partial{h_{t, u}^v}} + \frac{\partial{\mathcal{L}_\text{RNNT}}}{\partial{P(y_{u+1}| t, u)}} \frac{\partial{P(y_{u+1}| t, u)}}{\partial{h_{t, u}^v}}
\end{equation}
Next, consider the second part of each term in the summation,  $\frac{\partial{P(.| t, u)}}{\partial{h_{t, u}^v}}$. The gradients of the softmax  $y = \text{softmax}(x)$ are:
% i.e.
% $     y_i = \frac{\exp(x_i)}{\sum_j \exp(x_j)} $. So we have:
\begin{equation}
    \frac{\partial y_i}{\partial x_j} = \begin{cases}
        -y_i y_j, i \neq j \\
        y_i (1 - y_i) , i = j \
    \end{cases}
\end{equation}
Now we could simplify Eq.~\ref{2terms} based on different $v$. When $v = \O$, we have
\begin{equation}
\label{2terms_blank}
\begin{aligned}
\frac{\partial{\mathcal{L}_\text{RNNT}}}{\partial{h_{t, u}^{\O}}}  = & \frac{\partial{\mathcal{L}_\text{RNNT}}}{\partial{P(\O| t, u)}} \frac{\partial{P(\O| t, u)}}{\partial{h_{t, u}^{\O}}} + \frac{\partial{\mathcal{L}}_\text{RNNT}}{\partial{P(y_{u+1}| t, u)}} \frac{\partial{P(y_{u+1}| t, u)}}{\partial{h_{t, u}^{\O}}} \\
 = & -\frac{\alpha(t, u)\beta{(t+1,u)}}{P_\text{RNNT}(y|x)} P(\O| t, u) (1 - P(\O| t, u))  +\frac{\alpha(t, u)\beta{(t,u+1)}}{P_\text{RNNT}(y|x)} P(y_{u+1}| t, u)  P(\O| t, u) \\
= & \frac{P(\O|t, u) \alpha(t,u) }{P_\text{RNNT}(y|x)} \biggl[\underbrace{\beta(t+1, u) P(\O| t, u)  + \beta(t, u+1) P(y_{u+1} | t, u)}_{\beta(t, u)} -\beta(t+1, u) \biggr] \\
= & \frac{P(\O|t, u) \alpha(t,u) }{P_\text{RNNT}(y|x)} \biggl[\beta(t, u)-\beta(t+1, u)  \biggr] \\
\end{aligned}
\end{equation}
Similarly, when $v = y_{u+1}$, we have
\begin{equation}
\label{2terms_y_u_plus_1}
\begin{aligned}
\frac{\partial{\mathcal{L}_\text{RNNT}}}{\partial{h_{t, u}^{y_{u+1}}}}  = & \frac{\partial{\mathcal{L}_\text{RNNT}}}{\partial{P(\O| t, u)}} \frac{\partial{P(\O| t, u)}}{\partial{h_{t, u}^{y_{u+1}}}} + \frac{\partial{\mathcal{L}_\text{RNNT}}}{\partial{P(y_{u+1}| t, u)}} \frac{\partial{P(y_{u+1}| t, u)}}{\partial{h_{t, u}^{y_{u+1}}}} \\
 = & \frac{\alpha(t, u)\beta{(t+1,u)}}{P_\text{RNNT}(y|x)} P(\O| t, u)  P(y_{u+1}| t, u)  -\frac{\alpha(t, u)\beta{(t,u+1)}}{P_\text{RNNT}(y|x)} P(y_{u+1}| t, u)  (1 - P(y_{u+1}| t, u)) \\
= & \frac{P(y_{u+1}|t, u) \alpha(t,u) }{P_\text{RNNT}(y|x)} \biggl[\underbrace{\beta(t+1, u) P(\O|t,u)+ \beta(t, u+1) P(y_{u+1} | t, u)}_{\beta(t, u)} -\beta(t, u+1)   \biggr] \\
= & \frac{P(y_{u+1}|t, u) \alpha(t,u) }{P_\text{RNNT}(y|x)} \biggl[\beta(t, u)-\beta(t, u+1)  \biggr] 
\end{aligned}
\end{equation}
Lastly, when $v \neq \O$ and $v \neq y_{u+1}$, we have
\begin{equation}
\label{2terms_other}
\begin{aligned}
\frac{\partial{\mathcal{L}_\text{RNNT}}}{\partial{h_{t, u}^{v}}}  = & \frac{\partial{\mathcal{L}_\text{RNNT}}}{\partial{P(\O| t, u)}} \frac{\partial{P(\O| t, u)}}{\partial{h_{t, u}^{v}}} + \frac{\partial{\mathcal{L}_\text{RNNT}}}{\partial{P(y_{u+1}| t, u)}} \frac{\partial{P(y_{u+1}| t, u)}}{\partial{h_{t, u}^{v}}} \\
 = & \frac{\alpha(t, u)\beta{(t+1,u)}}{P_\text{RNNT}(y|x)} P(\O| t, u)  P(v| t, u)  +\frac{\alpha(t, u)\beta{(t,u+1)}}{P_\text{RNNT}(y|x)} P(y_{u+1}| t, u)   P(v| t, u) \\
= & \frac{P(v|t, u) \alpha(t,u) }{P_\text{RNNT}(y|x)} \biggl[\underbrace{\beta(t+1, u) P(\O|t,u)+ \beta(t, u+1) P(y_{u+1} | t, u)}_{\beta(t, u)}   \biggr] \\
= & \frac{P(v|t, u) \alpha(t,u) \beta(t, u) }{P_\text{RNNT}(y|x)}  
\end{aligned}
\end{equation}
Combining Eq. \ref{2terms_blank}, \ref{2terms_y_u_plus_1}, \ref{2terms_other}, we get the gradients of Transducer loss over pre-softmax logits, shown in Eq.~\ref{rnnt_func_merging}:
\begin{equation}
\label{rnnt_func_merging}
    \frac{\partial{\mathcal{L}_\text{RNNT}}}{\partial h^v_{t,u}} = \frac{P(v|t, u) \alpha(t, u)}{P_\text{RNNT}(y | x)}\Biggl[ \beta(t, u) - \begin{cases}
    \beta(t, u+1), &  v = y_{u + 1} \\
    \beta(t + 1, u), &  v = \O \\
    0, & \text{otherwise}
    \end{cases}
    \Biggr]
\end{equation}

\subsection{Softmax function merging for TDT}
The derivation of pre-softmax logit gradients for TDT loss is slightly more complex than the conventional Transducer but follows similar steps. First we follow the chain rule by writing the gradient as the summation of terms, and then listing only the  non-zero terms in the sum,
\begin{align}
\label{tdt_2_terms}
\begin{aligned}
    \frac{\partial{\mathcal{L}_\text{TDT}}}{\partial{h_{t, u}^v}} &= \sum_{v \in \mathcal{V}  } \frac{\partial{\mathcal{L}_\text{TDT}}}{\partial{P(v| t, u)}} \frac{\partial{P(v| t, u)}}{\partial{h_{t, u}^v}}  \\
    & = \frac{\partial{\mathcal{L}_\text{TDT}}}{\partial{P(\O| t, u)}} \frac{\partial{P(\O| t, u)}}{\partial{h_{t, u}^v}} + \frac{\partial{\mathcal{L}_\text{TDT}}}{\partial{P(y_{u+1}| t, u)}} \frac{\partial{P(y_{u+1}| t, u)}}{\partial{h_{t, u}^v}}
\end{aligned}
\end{align}
Now we simplify this depending on different $v$. 
\subsubsection{The case when $v=\O$}
\begin{equation}
\begin{aligned}
 \frac{\partial{\mathcal{L}_\text{TDT}}}{\partial{h_{t, u}^{\O}}} = & \frac{\partial{\mathcal{L}_\text{TDT}}}{\partial{P(\O| t, u)}} \frac{\partial{P(\O| t, u)}}{\partial{h_{t, u}^{\O}}} + \frac{\partial{\mathcal{L}_\text{TDT}}}{\partial{P(y_{u+1}| t, u)}} \frac{\partial{P(y_{u+1}| t, u)}}{\partial{h_{t, u}^{\O}}}  \\
 = & -\frac{\alpha(t, u) \sum_{d \in \mathcal{D} \setminus \{0\}} {\beta(t + d,u) P_D(d|t, u)}} {P_\text{TDT}(y|x)} P(\O| t, u) (1 - P(\O| t, u)) \\
 & + \frac{\alpha(t, u) \sum_{d \in \mathcal{D}}                 {\beta(t + d,u+1)P_D(d|t,u)}} {P_\text{TDT}(y|x)} P(\O| t, u) P(y_{u+1}|t, u) \\
 = & \frac{P(\O|t, u) \alpha(t, u)}{P_\text{TDT}(y|x)}\\
 & \Biggl[ ( P(\O|t, u) - 1) \sum_{d \in \mathcal{D} \setminus \{0\}} {\beta(t + d,u) P_D(d|t, u)} + P(y_{u+1}|t, u) \sum_{d \in \mathcal{D}}                 {\beta(t + d,u+1)P_D(d|t,u)}  \Biggr]
 \end{aligned} 
\end{equation}
The term in the bracket could be further simplified as,
\begin{equation}\label{tdt_blank_2}
\begin{aligned}
& ( P(\O|t, u) - 1) \sum_{d \in \mathcal{D} \setminus \{0\}} {\beta(t + d,u) P_D(d|t, u)} + P(y_{u+1}|t, u) \sum_{d \in \mathcal{D}}                 {\beta(t + d,u+1)P_D(d|t,u)}   \\
= & -\sum_{d \in \mathcal{D} \setminus \{0\}} {\beta(t + d,u) P_D(d|t, u)} \\
& + P(\O|t, u) \sum_{d \in \mathcal{D} \setminus \{0\}} {\beta(t + d,u) P_D(d|t, u)} + P(y_{u+1}|t, u) \sum_{d \in \mathcal{D}}                 {\beta(t + d,u+1)P_D(d|t,u)}  \\
= & -\sum_{d \in \mathcal{D} \setminus \{0\}} {\beta(t + d,u) P_D(d|t, u)} + \Biggl[ \underbrace{\sum_{d \in \mathcal{D} \setminus \{0\}} {\beta(t + d,u) P(\O, d|t, u)} + \sum_{d \in \mathcal{D}}{\beta(t + d,u+1)P(y_{u+1}, d|t,u)}}_{\beta(t, u)} \Biggr] \\
= &\beta(t, u) -\sum_{d \in \mathcal{D} \setminus \{0\}} {\beta(t + d,u) P_D(d|t, u)} 
 \end{aligned}
\end{equation}
Apply the results of \ref{tdt_blank_2} to \ref{tdt_2_terms}, we get
\begin{equation}\label{tdt_fm_1}
\frac{\partial{\mathcal{L}_\text{TDT}}}{\partial{h_{t, u}^{\O}}} =    \frac{P(\O|t, u) \alpha(t, u)}{P_\text{TDT}(y|x)} \Biggl[\beta(t, u) -\sum_{d \in \mathcal{D} \setminus \{0\}} {\beta(t + d,u) P_D(d|t, u)}  \Biggr]
\end{equation}

% where $\mathcal{V}$ represents a set of all vocabulary including the $\O$ token.
% In the summation, $\frac{\partial{\mathcal{L}}}{\partial{P(v| t, u)}}$ can be calculated via Eqn \ref{rnnt_grad}, and although this summation is over all elements in $\mathcal{V}$, only two of them are non-zero - the one where $v' = \O$ and where $v' = y_{u+1}$. Therefore, we could simply Equation \ref{function_merging_all} as, 

% \begin{equation}
% \label{2terms}
% \frac{\partial{\mathcal{L}_\text{transducer}}}{\partial{h_{t, u}^v}} = \frac{\partial{\mathcal{L}}}{\partial{P(\O| t, u)}} \frac{\partial{P(\O| t, u)}}{\partial{h_{t, u}^v}} + \frac{\partial{\mathcal{L}_\text{transducer}}}{\partial{P(y_{u+1}| t, u)}} \frac{\partial{P(y_{u+1}| t, u)}}{\partial{h_{t, u}^v}}
% \end{equation}

% \begin{align}
%     % \frac{\partial{L}}{\partial{h_{t, u}^v}} &= \frac{P_T(v| t, u) \: \alpha(t,u)}{P(y^*|x)} \left[ \beta(t, u) - \sum_{v' \in v} B(v', t, u)\right] 
%     \frac{\partial{L}}{\partial{h_{t, u}^v}} &= -\frac{\alpha(t,u) \: B(v, t, u) }{P(y^*|x)} \frac{\partial{P_T(v| t, u)}}{\partial{h_{t, u}^v}} \\
%     &= -\frac{\alpha(t,u) B(v, t, u) }{P(y^*|x)} P_T(v| t, u) \left( 1 - \frac{\beta(t, u)}{P(y^*|x)}  \right) \\
%     &= \frac{P_T(v| t, u) \: \alpha(t,u)}{P(y^*|x)} \left[ \beta(t, u) - B(v, t, u)\right] 
% \end{align}

\subsubsection{The case when $v=y_{u+1}$}
\begin{equation}
\begin{aligned}
 \frac{\partial{\mathcal{L}_\text{TDT}}}{\partial{h_{t, u}^{y_{u+1}}}} = & \frac{\partial{\mathcal{L}_\text{TDT}}}{\partial{P(\O| t, u)}} \frac{\partial{P(\O| t, u)}}{\partial{h_{t, u}^{y_{u+1}}}} + \frac{\partial{\mathcal{L}_\text{TDT}}}{\partial{P(y_{u+1}| t, u)}} \frac{\partial{P(y_{u+1}| t, u)}}{\partial{h_{t, u}^{y_{u+1}}}}  \\
 = & \frac{\alpha(t, u) \sum_{d \in \mathcal{D} \setminus \{0\}} {\beta(t + d,u) P_D(d|t, u)}} {P_\text{TDT}(y|x)} P(\O| t, u) P(y_{u+1}| t, u) \\
 & + \frac{\alpha(t, u) \sum_{d \in \mathcal{D}}                 {\beta(t + d,u+1)P_D(d|t,u)}} {P_\text{TDT}(y|x)} P(y_{u+1}| t, u) (P(y_{u+1}|t, u) - 1) \\
 = & \frac{P(y_{u+1}|t, u) \alpha(t, u)}{P_\text{TDT}(y|x)}\\
 & \Biggl[ P(\O|t, u)  \sum_{d \in \mathcal{D} \setminus \{0\}} {\beta(t + d,u) P_D(d|t, u)} + (P(y_{u+1}|t, u) - 1) \sum_{d \in \mathcal{D}}                 {\beta(t + d,u+1)P_D(d|t,u)}  \Biggr]
 \end{aligned} 
\end{equation}
The term in the bracket could be further simplified as,
\begin{equation}\label{tdt_y_2}
\begin{aligned}
&  P(\O|t, u) \sum_{d \in \mathcal{D} \setminus \{0\}} {\beta(t + d,u) P_D(d|t, u)} + (P(y_{u+1}|t, u) - 1) \sum_{d \in \mathcal{D}}                 {\beta(t + d,u+1)P_D(d|t,u)}   \\
= & -\sum_{d \in \mathcal{D}} {\beta(t + d,u+1) P_D(d|t, u)} + P(\O|t, u) \sum_{d \in \mathcal{D} \setminus \{0\}} {\beta(t + d,u) P_D(d|t, u)} \\
& + P(y_{u+1}|t, u) \sum_{d \in \mathcal{D}}                 {\beta(t + d,u+1)P_D(d|t,u)}  \\
= & -\sum_{d \in \mathcal{D} } {\beta(t + d,u+1) P_D(d|t, u)} + \Biggl[ \underbrace{\sum_{d \in \mathcal{D} \setminus \{0\}} {\beta(t + d,u) P(\O, d|t, u)} + \sum_{d \in \mathcal{D}}{\beta(t + d,u+1)P(y_{u+1}, d|t,u)}}_{\beta(t, u)} \Biggr] \\
= &\beta(t, u) -\sum_{d \in \mathcal{D} } {\beta(t + d,u+1) P_D(d|t, u)} 
 \end{aligned}
\end{equation}
Apply the results of \ref{tdt_y_2} to \ref{tdt_2_terms}, we get
\begin{equation}\label{tdt_fm_2}
\frac{\partial{\mathcal{L}_\text{TDT}}}{\partial{h_{t, u}^{y_{u+1}}}} =    \frac{P(y_{u+1}|t, u) \alpha(t, u)}{P_\text{TDT}(y|x)} \Biggl[\beta(t, u) -\sum_{d \in \mathcal{D} } {\beta(t + d,u + 1) P_D(d|t, u)}  \Biggr]
\end{equation}

\subsubsection{The case when $v\neq y_{u+1}$ and $v \neq \O$}
\begin{equation}
\begin{aligned}
 \frac{\partial{\mathcal{L}_\text{TDT}}}{\partial{h_{t, u}^{v}}} = & \frac{\partial{\mathcal{L}_\text{TDT}}}{\partial{P(\O| t, u)}} \frac{\partial{P(\O| t, u)}}{\partial{h_{t, u}^{v}}} + \frac{\partial{\mathcal{L}_\text{TDT}}}{\partial{P(y_{u+1}| t, u)}} \frac{\partial{P(y_{u+1}| t, u)}}{\partial{h_{t, u}^{v}}}  \\
 = & \frac{\alpha(t, u) \sum_{d \in \mathcal{D} \setminus \{0\}} {\beta(t + d,u) P_D(d|t, u)}} {P_\text{TDT}(y|x)} P(\O| t, u) P(y_{u+1}| t, u) \\
 & + \frac{\alpha(t, u) \sum_{d \in \mathcal{D}}                 {\beta(t + d,u+1)P_D(d|t,u)}} {P_\text{TDT}(y|x)} P(y_{u+1}| t, u) P(y_{u+1}|t, u) \\
 = & \frac{P(y_{u+1}|t, u) \alpha(t, u)}{P_\text{TDT}(y|x)}\\
 & \Biggl[ P(\O|t, u)  \sum_{d \in \mathcal{D} \setminus \{0\}} {\beta(t + d,u) P_D(d|t, u)} + P(y_{u+1}|t, u)  \sum_{d \in \mathcal{D}}                 {\beta(t + d,u+1)P_D(d|t,u)}  \Biggr]
 \end{aligned} 
\end{equation}

The term in the bracket could be further simplified as,
\begin{equation}\label{tdt_other_2}
\begin{aligned}
&  P(\O|t, u) \sum_{d \in \mathcal{D} \setminus \{0\}} {\beta(t + d,u) P_D(d|t, u)} + P(y_{u+1}|t, u)  \sum_{d \in \mathcal{D}}                 {\beta(t + d,u+1)P_D(d|t,u)}   \\
= &  \sum_{d \in \mathcal{D} \setminus \{0\}} {\beta(t + d,u) P_D(d|t, u)} + P(y_{u+1}|t, u) \sum_{d \in \mathcal{D}}                 {\beta(t + d,u+1)P_D(d|t,u)}  \\
= &  \sum_{d \in \mathcal{D} \setminus \{0\}} {\beta(t + d,u) P(\O, d|t, u)} + \sum_{d \in \mathcal{D}}{\beta(t + d,u+1)P(y_{u+1}, d|t,u)} \\
= & \beta(t, u) 
 \end{aligned}
\end{equation}
Apply the results of \ref{tdt_other_2} to \ref{tdt_2_terms}, we get
\begin{equation}\label{tdt_fm_3}
\frac{\partial{\mathcal{L}_\text{TDT}}}{\partial{h_{t, u}^{v}}} =    \frac{P(v|t, u) \alpha(t, u) \beta(t, u)}{P_\text{TDT}(y|x)} 
\end{equation}
Combining the results from Equations \ref{tdt_fm_1}, \ref{tdt_fm_2}, \ref{tdt_fm_3}, we have
\begin{equation}
\frac{\partial{\mathcal{L}_\text{TDT}}}{\partial{h_{t, u}^{v}}} =    \frac{P(v|t, u) \alpha(t, u)}{P_\text{TDT}(y|x)}    \Biggl[ \beta(t, u) - \begin{cases}
\sum_{d \in \mathcal{D} \setminus \{0\}} {\beta(t + d,u) P_D(d|t, u)}, & v = \O \\
\sum_{d \in \mathcal{D} } {\beta(t + d,u + 1) P_D(d|t, u)},  & v = y_{u+1} \\
0, & \text{otherwise}
\end{cases}
\Biggr] 
\end{equation}

\section{Derivation of TDT Gradients with Logit Under-normalization}
In this section we derive the gradients of TDT with logit under-normalization introduced in Section \ref{sec:under_norm}). 
Let's denote the pseudo ``probability'' acquired with under-normalization, using the ``under-softmax'' operation of $y' = \frac{\text{softmax}(x)}{\exp(\sigma)}$ as $P'(v | t, u) = \frac{P(v | t, u)}{\exp(\sigma)}$. Let's first work out the gradients of the ``under-softmax'' operation. We assume $y = \text{softmax}(x)$ and thus $y' \exp(\sigma) = y$, then
\begin{equation}
\begin{aligned}
    \frac{\partial y'_i}{\partial x_j} = \frac{\partial y'_i}{\partial y_i} \frac{\partial y_i}{\partial x_j} & = \frac{1}{\exp(\sigma)} \begin{cases}
    y_i (1 - y_i), & i = j \\
    -y_i  y_j & i \neq j \\
    \end{cases} \\
    & = \frac{1}{\exp(\sigma)} \begin{cases}
    \exp(\sigma) y'_i (1 - \exp(\sigma) y'_i), & i = j \\
    -y'_i  y'_j \exp^2(\sigma)& i \neq j \\
    \end{cases} \\
    & = \begin{cases}
    y'_i (1 - \exp(\sigma) y'_i), & i = j \\
    -y'_i  y'_j \exp(\sigma)& i \neq j \\
    \end{cases} \\
\end{aligned}
\end{equation}
Next apply the chain rule:
\begin{align}
\label{sigma_chainrule}
\begin{aligned}
\frac{\partial{\mathcal{L}_\text{TDT}}}{\partial{h_{t, u}^v}} &= \sum_{v \in \mathcal{V}  } \frac{\partial{\mathcal{L}_\text{TDT}}}{\partial{P'(v| t, u)}} \frac{\partial{P'(v| t, u)}}{\partial{h_{t, u}^v}}  \\
    & = \frac{\partial{\mathcal{L}_\text{TDT}}}{\partial{P'(\O| t, u)}} \frac{\partial{P'(\O| t, u)}}{\partial{h_{t, u}^v}} + \frac{\partial{\mathcal{L}_\text{TDT}}}{\partial{P'(y_{u+1}| t, u)}} \frac{\partial{P'(y_{u+1}| t, u)}}{\partial{h_{t, u}^v}}
\end{aligned}
\end{align}
We would like to emphasize that under-normalization is only applied to token logits, not duration logits. 
Now, let's use $P'(y|x)$ to denote the pseudo ``probability'' of the sequence, computed with $P'(v|t,u)$ throughout the forward-backward algorithm. Now we further simplify the terms. 

\subsection{When $v$ is neither $\O $ nor $y_{u+1}$}
\begin{equation} \label{under_norm_1}
\begin{aligned}
 \frac{\partial{\mathcal{L}_\text{TDT}}}{\partial{h_{t, u}^v}} = & \frac{\partial{\mathcal{L}_\text{TDT}}}{\partial{P'(\O| t, u)}} \frac{\partial{P'(\O| t, u)}}{\partial{h_{t, u}^v}} + \frac{\partial{\mathcal{L}_\text{TDT}}}{\partial{P'(y_{u+1}| t, u)}} \frac{\partial{P'(y_{u+1}| t, u)}}{\partial{h_{t, u}^v}} \\
=  & \frac{\alpha(t, u) \sum_{d \in \mathcal{D} \setminus \{0\}} {\beta(t + d,u) P_D(d|t, u)}} {P'(y|x)} P'(\O| t, u) P'(v| t, u) \exp(\sigma) \\
 & + \frac{\alpha(t, u) \sum_{d \in \mathcal{D}}                 {\beta(t + d,u+1)P_D(d|t,u)}} {P'(y|x)} P'(y_{u+1}| t, u) P'(v|t, u) \exp(\sigma) \\
 = & \frac{\exp(\sigma) P'(v|t, u) \alpha(t, u)}{P'(y|x)}\\
 & \Biggl[\underbrace{ P'(\O|t, u)  \sum_{d \in \mathcal{D} \setminus \{0\}} {\beta(t + d,u) P_D(d|t, u)} + P'(y_{u+1}|t, u)  \sum_{d \in \mathcal{D}}                 {\beta(t + d,u+1)P_D(d|t,u)} }_{\beta(t,u)} \Biggr] \\
 = & \frac{\exp(\sigma) P'(v|t, u) \alpha(t, u) \beta(t,u)}{P'(y|x)} \\
 = & \frac{P(v|t, u) \alpha(t, u) \beta(t,u)}{P'(y|x)} \\
\end{aligned}
\end{equation}

\subsection{When $v = \O$}

\begin{equation} \label{under_norm_2}
\begin{aligned}
 \frac{\partial{\mathcal{L}_\text{TDT}}}{\partial{h_{t, u}^{\O}}} = & \frac{\partial{\mathcal{L}_\text{TDT}}}{\partial{P'(\O| t, u)}} \frac{\partial{P'(\O| t, u)}}{\partial{h_{t, u}^{\O}}} + \frac{\partial{\mathcal{L}_\text{TDT}}}{\partial{P'(y_{u+1}| t, u)}} \frac{\partial{P'(y_{u+1}| t, u)}}{\partial{h_{t, u}^{\O}}} \\
=  & \frac{\alpha(t, u) \sum_{d \in \mathcal{D} \setminus \{0\}} {\beta(t + d,u) P_D(d|t, u)}} {P'(y|x)} P'(\O| t, u)(P'(\O| t, u) \exp(\sigma) - 1)\\
 & + \frac{\alpha(t, u) \sum_{d \in \mathcal{D}}                 {\beta(t + d,u+1)P_D(d|t,u)}} {P'(y|x)} P'(y_{u+1}| t, u) P'(\O|t, u) \exp(\sigma) \\
 = & \frac{\exp(\sigma) P'(v|t, u) \alpha(t, u)}{P'(y|x)} \Biggl[ - \frac{1}{\exp(\sigma)} \sum_{d \in \mathcal{D} \setminus \{0\}} {\beta(t + d,u) P_D(d|t, u)}\\
 & + \underbrace{ P'(\O|t, u)  \sum_{d \in \mathcal{D} \setminus \{0\}} {\beta(t + d,u) P_D(d|t, u)} + P'(y_{u+1}|t, u)  \sum_{d \in \mathcal{D}}                 {\beta(t + d,u+1)P_D(d|t,u)} }_{\beta(t,u)}  \Biggr] \\
 = & \frac{\exp(\sigma) P'(v|t, u) \alpha(t, u) }{P'(y|x)} \bigg[ \beta(t,u) -\frac{1}{\exp(\sigma)}  \sum_{d \in \mathcal{D} \setminus \{0\}} {\beta(t + d,u) P_D(d|t, u)} \bigg] \\
 = & \frac{P(v|t, u) \alpha(t, u)}{P'(y|x)} \bigg[ \beta(t,u) - \frac{1}{\exp(\sigma)}  \sum_{d \in \mathcal{D} \setminus \{0\}} {\beta(t + d,u) P_D(d|t, u)} \bigg] \\
\end{aligned}
\end{equation}

\subsection{When $v = y_{u+1}$}

\begin{equation} \label{under_norm_3}
\begin{aligned}
 \frac{\partial{\mathcal{L}_\text{TDT}}}{\partial{h_{t, u}^{y_{u+1}}}} = & \frac{\partial{\mathcal{L}_\text{TDT}}}{\partial{P'(\O| t, u)}} \frac{\partial{P'(\O| t, u)}}{\partial{h_{t, u}^{y_{u+1}}}} + \frac{\partial{\mathcal{L}_\text{TDT}}}{\partial{P'(y_{u+1}| t, u)}} \frac{\partial{P'(y_{u+1}| t, u)}}{\partial{h_{t, u}^{y_{u+1}}}} \\
=  & \frac{\alpha(t, u) \sum_{d \in \mathcal{D} \setminus \{0\}} {\beta(t + d,u) P_D(d|t, u)}} {P'(y|x)} P'(y_{u+1}| t, u)P'(\O| t, u) \exp(\sigma)\\
 & + \frac{\alpha(t, u) \sum_{d \in \mathcal{D}}                 {\beta(t + d,u+1)P_D(d|t,u)}} {P'(y|x)} P'(y_{u+1}| t, u) (P'(y_{u+1}|t, u) \exp(\sigma) - 1) \\
 = & \frac{\exp(\sigma) P'(v|t, u) \alpha(t, u)}{P'(y|x)} \Biggl[ - \frac{1}{\exp(\sigma)} \sum_{d \in \mathcal{D} } {\beta(t + d,u+1) P_D(d|t, u)}\\
 & + \underbrace{ P'(\O|t, u)  \sum_{d \in \mathcal{D} \setminus \{0\}} {\beta(t + d,u) P_D(d|t, u)} + P'(y_{u+1}|t, u)  \sum_{d \in \mathcal{D}}                 {\beta(t + d,u+1)P_D(d|t,u)} }_{\beta(t,u)}  \Biggr] \\
 = & \frac{\exp(\sigma) P'(v|t, u) \alpha(t, u) }{P'(y|x)} \bigg[ \beta(t,u) -\frac{1}{\exp(\sigma)}  \sum_{d \in \mathcal{D} } {\beta(t + d,u+1) P_D(d|t, u)} \bigg] \\
 = & \frac{P(v|t, u) \alpha(t, u)}{P'(y|x)} \bigg[ \beta(t,u) - \frac{1}{\exp(\sigma)}  \sum_{d \in \mathcal{D}} {\beta(t + d,u+1) P_D(d|t, u)} \bigg] \\
\end{aligned}
\end{equation}
Combining Eq. \ref{under_norm_1}, \ref{under_norm_2} and \ref{under_norm_3}, we have the TDT gradients with under-normalization as,
\begin{equation}
\frac{\partial{\mathcal{L}_\text{TDT}}}{\partial{h_{t, u}^{v}}} =  \frac{P(v|t, u) \alpha(t, u)}{P'(y|x)} \bigg[ \beta(t,u) - \frac{1}{\exp(\sigma)} \begin{cases}
 \sum_{d \in \mathcal{D}} {\beta(t + d,u+1) P_D(d|t, u)}, & v = y_{u+1} \\
 \sum_{d \in \mathcal{D}\setminus \{0\}} {\beta(t + d,u) P_D(d|t, u)}, & v = \O \\
0, & \text{otherwise} 
\end{cases}
\bigg]
\end{equation}

\section{Analysis of Token-and-Duration Transducer Alignments}

TDT models are capable of learning the alignment $P(\mathcal{A}_{t, u} | x)$ between an input sequence ($S$) and the corresponding target sequence ($S'$). This quantity represents the total probability mass that goes through the state $(t, u)$ in the lattice.
We use the definition of alignment $P(\mathcal{A}_{t, u} | x)$ from \cite{yu2021fastemit},  computed as,
\begin{equation}
    P(\mathcal{A}_{t, u} | x) = \alpha(t, u) \beta(t, u)
\end{equation}
In the following section, we construct a set of force-alignment experiments to determine the effect of input-output sequence interactions and hyper-parameter choices. 
Unless stated otherwise, all following experiments are with a simulated joint tensor ($J_S \in R^{T \times U \times (V + 1 + N_d)}$) for some input sequence ($S$) sampled from the normal distribution ($N(0, 1)$) and a target sequence ($S \in \mathbb{Z}^{U}$) generated from the discrete uniform distribution ($U\{1, \ldots, V \}$), where $N_d$ refers the the the number of duration tokens and $V$ is the size of the vocabulary of the token set. We use the PyTorch  \cite{paszke_pytorch} to optimize $J_S$ using $S'$ as the target sequence. 
% or \ref{eqn:TDT_loss_fastemit} depending on whether we use FastEmit or not respectively. 
We minimize he TDT loss defined in Eq.~\ref{eqn:TDT_loss} using the Adam optimizer \cite{kingma_adam} with 
a fixed learning rate of 0.1 
for 100 update steps. Once $J_S$ is optimized, we compute 
 $P(\mathcal{A}_{t, u} | x)$ and plot the $T \times U$ alignment matrix. We use a fixed random seed so as to reproduce the same $J_S$ and $S'$ when $T$, $U$ and $V$ are kept constant and chose a fixed value of $T = 70$, $U = 10$, $V = 5$ (chosen simply to speed up convergence), number of duration tokens ($N_d = 8$), $\sigma = 0.05$, $\omega = 0.0$ and $\text{fastemit } \lambda = 0.0$ for the experiments unless explicitly mentioned. 

\subsection{Effect of Durations on TDT Alignments}
\label{sec:durations_tdt}

\begin{figure} [t]
    \centering
    \begin{minipage}{0.45\textwidth}
        \centering
        \includegraphics[width=1.0\textwidth]{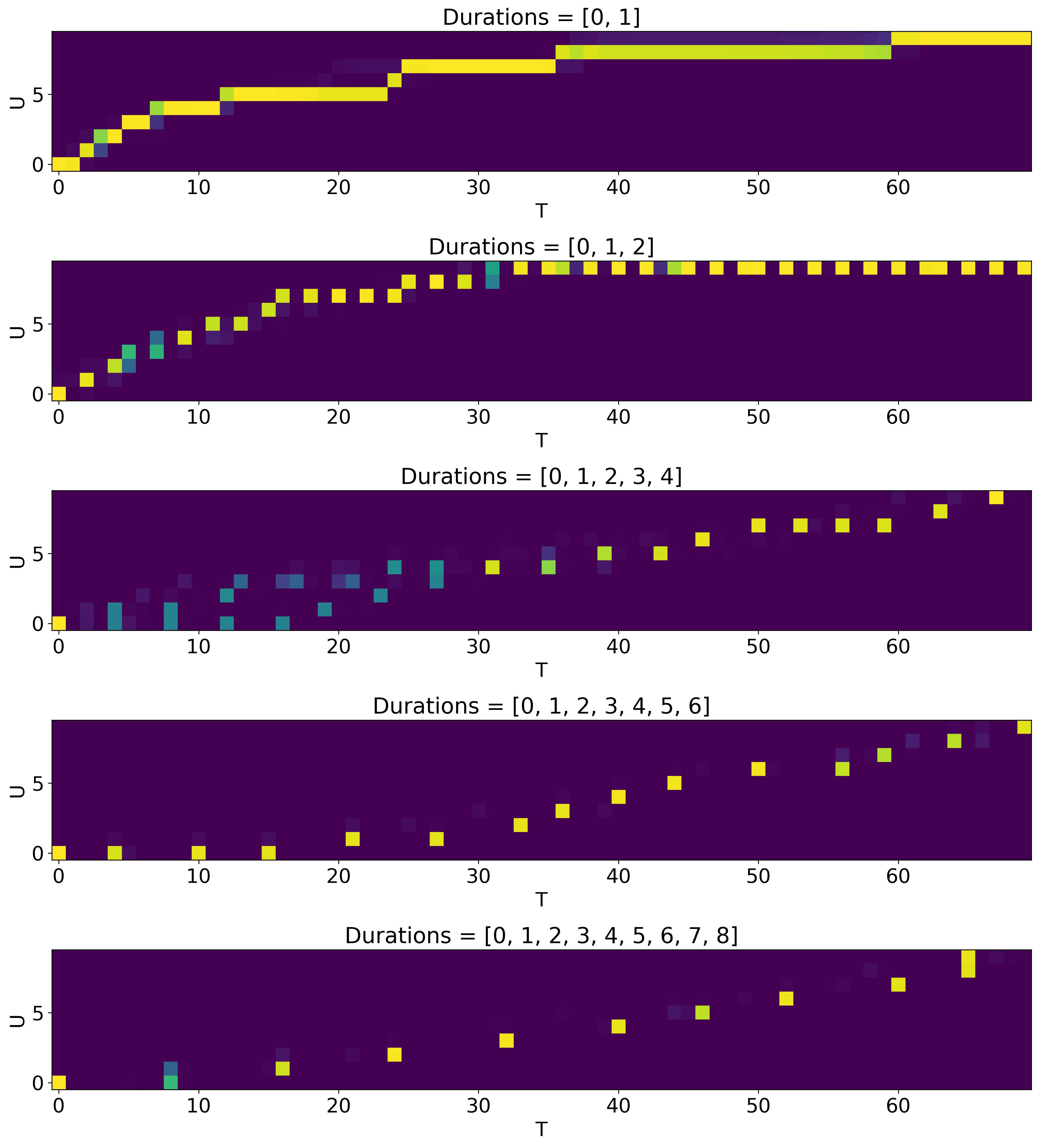}
        \caption{Alignment ($P(\mathcal{A}_{t, u} | x)$ as a function of the duration ($\mathcal{D}$) supported by the TDT model. Simulated joint ($J_S$) trained with a larger number of duration tokens ($N_d$) possess alignments with correspondingly longer gaps between time steps ($t \in T$) for each token prediction ($u \in U$)}
       \label{fig:appendix_durations}
    \end{minipage}\hfill
    \begin{minipage}{0.45\textwidth}
        \centering
        \includegraphics[width=1.0\textwidth]{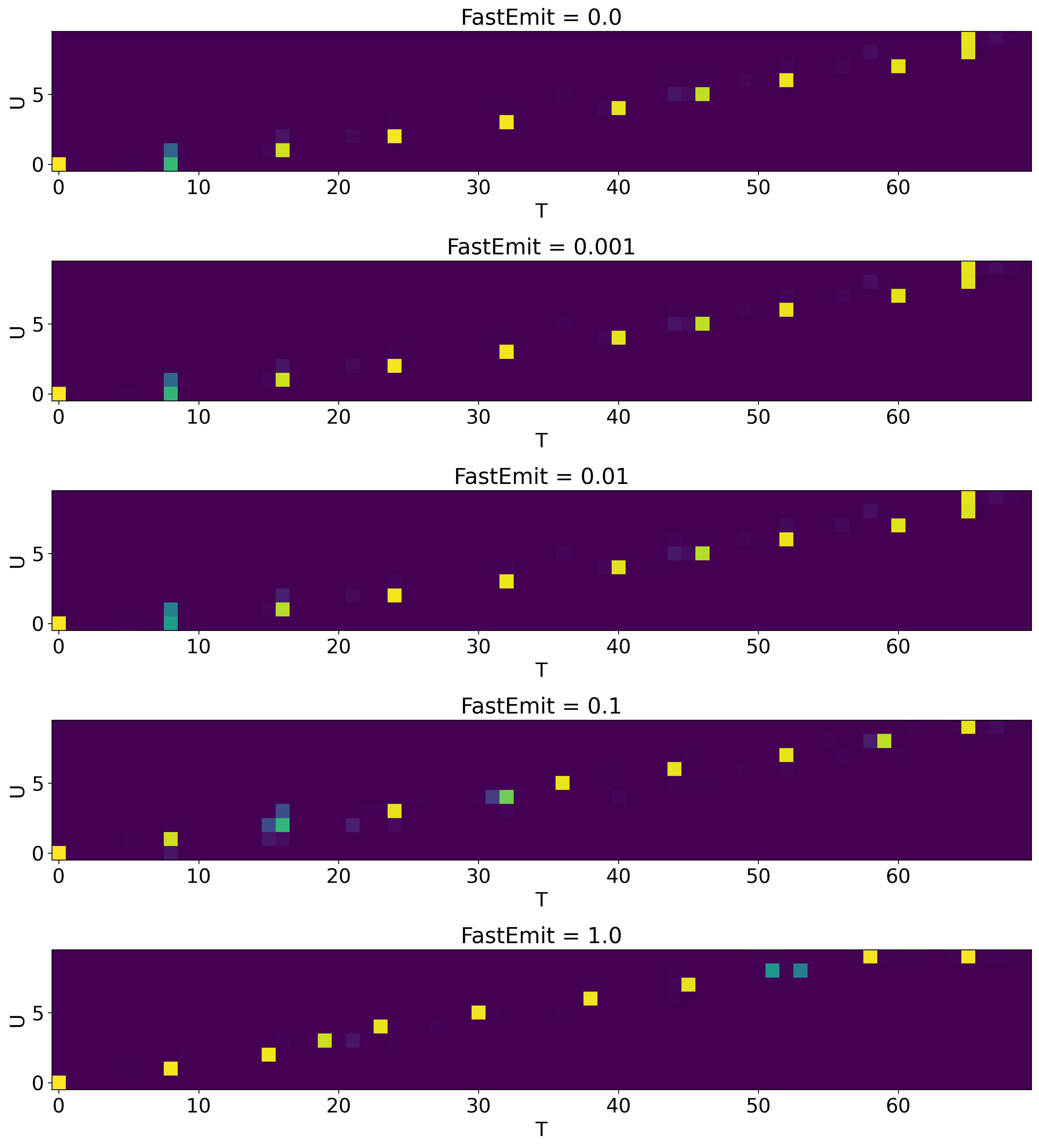} 
        \caption{Alignment ($P(\mathcal{A}_{t, u} | x)$ as a function of the FastEmit regularization strength ($\lambda$). Simulated joint ($J_S$) trained with varying $\lambda$ possess alignments with a correspondingly shorter delay between each token prediction ($u \in U$) as compared to the baseline of $\lambda = 0$}
        \label{fig:appendix_fastemit}
    \end{minipage}
\end{figure}

\iffalse {
\begin{figure}[b]
\centering
    \includegraphics[width=0.5\textwidth]{appendix_durations.jpg}
    \caption{Alignment ($P(\mathcal{A}_{t, u} | x)$ as a function of the duration ($\mathcal{D}$) supported by the TDT model. Simulated joint ($J_S$) trained with a larger number of duration tokens ($N_d$) possess alignments with correspondingly longer gaps between time steps ($t \in T$) for each token prediction ($u \in U$)}
    \label{fig:appendix_durations}
\end{figure}

\begin{figure}[h!]
    \centering
    \includegraphics[width=0.55\textwidth]{appendix_fastemit.jpg}
    \caption{Alignment ($P(\mathcal{A}_{t, u} | x)$ as a function of the FastEmit regularization strength ($\lambda$). Simulated joint ($J_S$) trained with varying $\lambda$ possess alignments with a correspondingly shorter delay between each token prediction ($u \in U$) as compared to the baseline of $\lambda = 0$}
    \label{fig:appendix_fastemit}
\end{figure}
} \fi

In a TDT model, one head emits the token while another predicts the duration of the token (say $\mathcal{D} \in \{ 0, 1, \ldots, (N_d)-1 \}$) where $N_d$ is the number of duration tokens. In the following set of experiments, we attempt to optimize $J_S$  while modifying only the duration set. Given that $T >> U$, and $\sigma > 0$, we expect the alignment to intuitively contain longer-duration tokens as $N_d$ becomes larger.

In Fig.~\ref{fig:appendix_durations}, we see that the learned alignment precisely matches our expectations. As the $N_d$ grows larger, the model selects longer durations, significantly reducing the number of decoding steps required. A natural effect of selecting longer tokens is that token emissions are significantly delayed compared to the baseline of $\mathcal{D} \in \{ 0, 1 \}$ which can be considered an approximation of conventional Transducer alignment with single duration step per token emitted. 

Note, that longer duration tokens are enabled by the large difference in the input and target sequence lengths ($T = 70; \: U = 10$), reducing to a $\frac{T}{U} = 7:1$ ratio. This ratio of sequence lengths is slightly larger than the observed ratio of acoustic sequence length versus the corresponding sub-word encoded target text tokens in Librispeech with a sufficiently large sub-word encoding vocabulary, close to $\frac{T_\text{LS}}{U_\text{LS}} \approx 5.5:1$. We expect that target token encoding schemes such as character-based encoding will diminish this ratio of sequence lengths to approximately $\frac{T}{U} \approx 2:1$, thereby preventing long-duration tokens from being emitted frequently. 

\vspace{-4pt}
\subsection{Effect of FastEmit on Alignments}
\label{sec:fastemit_tdt}
One reduce the delay of token emission using FastEmit method proposed in \cite{yu2021fastemit}. 
As can be seen in Section \ref{sec:durations_tdt}, when the number of duration tokens ($N_d$) is large, the emission of tokens ($u \in U$) is delayed significantly which may hinder latency-sensitive applications. In the following section, we discuss the utilization of FastEmit \cite{yu2021fastemit} as a strong regularization scheme in order to prevent the model from deferring token emission to such a degree.
FastEmit introduces a hyperparameter $\lambda$ and scales the gradients to token probabilities by $1 + \lambda$ and keeping the gradients to blank probabilities unchanged. We attempt to simply train the same simulated joint ($J_S$) optimized with different strengths of the $\lambda$ scaling factor for FastEmit.

In Fig. \ref{fig:appendix_fastemit}, we find that the effect of FastEmit regularization strength constant ($\lambda$) has a substantial effect on reducing the delay between token emissions. It must be noted that $\lambda > 1e\textit{-}2$ is not realistically applicable when training on non-synthetic data, as the strength of the regularization term will cause the model to diverge, but in this simulated setting, it has been done in order to explicitly show the effect of FastEmit on token emission delay.

%\vspace{-6pt}
An important observation is that when comparing the alignment of Figure \ref{fig:appendix_durations} (1st row) and Figure \ref{fig:appendix_fastemit} (5th row), the first 5 $\sim$ 6 token emissions occur rapidly with the duration set $N_d = 1$ but are delayed by a significant number of steps for $N_d = 8$. FastEmit does improve the token emission of the first few tokens, however since each token presents a duration of roughly 8 timesteps, the overall latency is significantly higher. This can be tackled by carefully modifying the strength of $\sigma$ along with $\lambda$, so as to discourage long tokens while encouraging faster emission of tokens if latency is a concern.

\subsection{Effect of Input Sequence length on Duration Prediction}
\label{sec:seq_len_tdt}

\begin{figure}[t]
    \centering
    \includegraphics[width=0.55\textwidth]{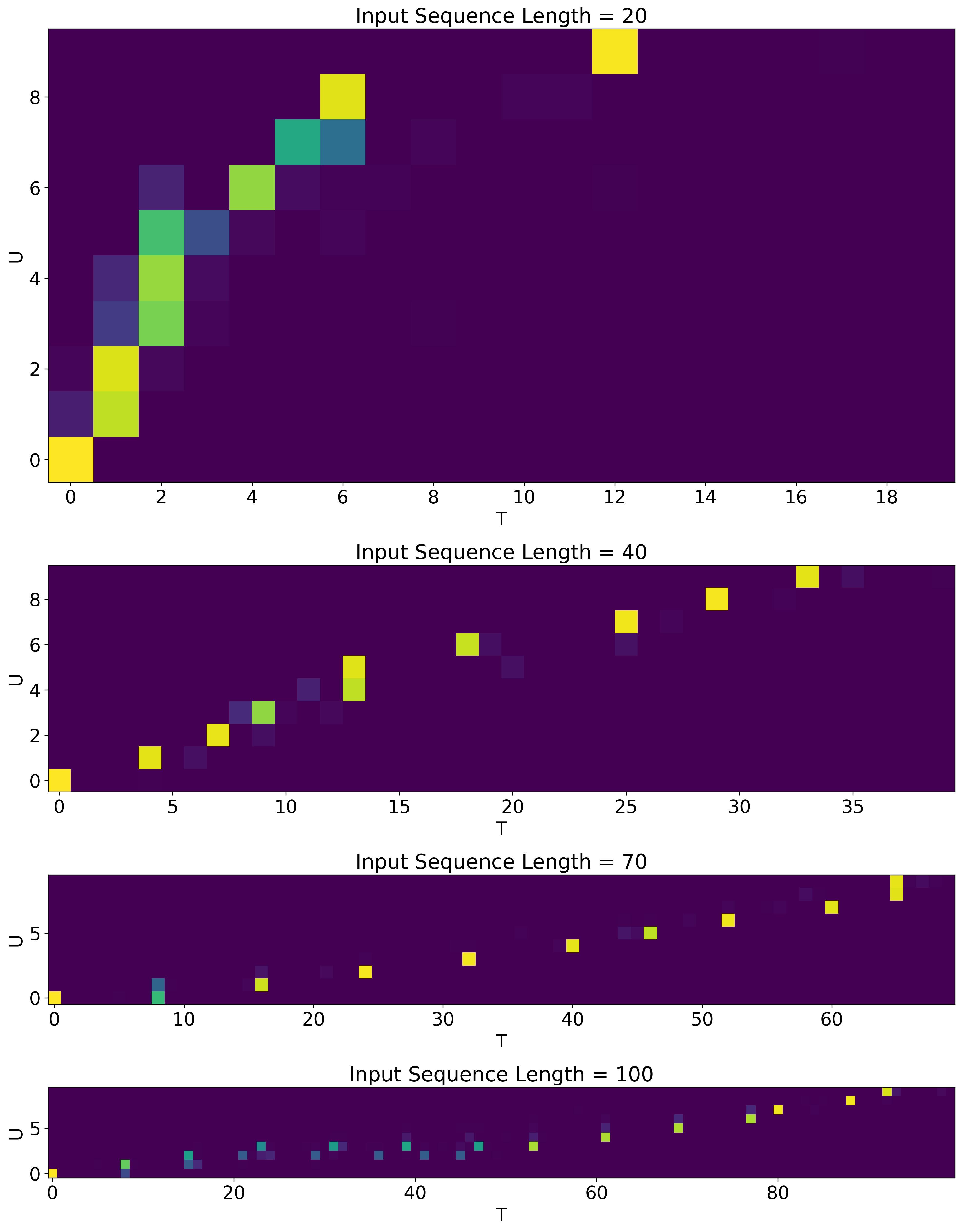}
    \caption{Alignment ($P(\mathcal{A}_{t, u} | x)$ as a function of the input sequence length ($T$). For each $T$, the corresponding simulated joint ($J_{S_T}$) is trained. The alignments for each $J_{S_T}$ show that as $T$ grows, thereby causing a larger ratio of $\frac{T}{U}$, the model begins to emit tokens with longer duration, as well as increased delay in token emission.}
    \label{fig:appendix_seq_len_non_fastemit}
\end{figure}
\begin{figure}[h!]
    \centering
    \includegraphics[width=0.55\textwidth]{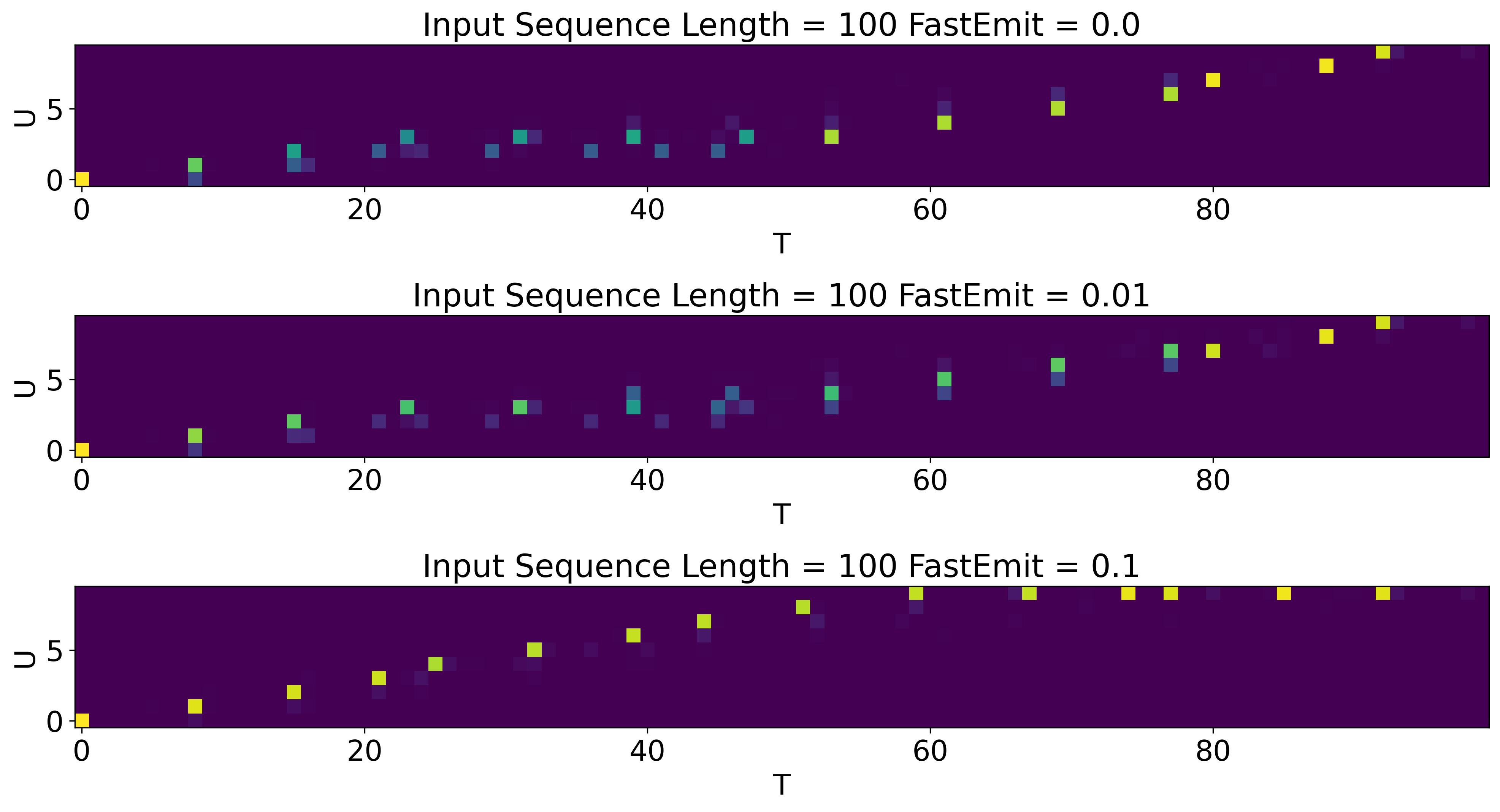}
    \caption{Alignment ($P(\mathcal{A}_{t, u} | x)$ as a function of the input sequence length ($T$), with FastEmit weight ($\lambda$) as regularization. For given $T >> U$, the corresponding simulated joint ($J_{S_T}$) is trained. The alignments for each $J_{S_T}$ show that as $T$ grows, thereby causing a larger ratio of $\frac{T}{U}$, the model begins to emit tokens with longer duration, as well as increased delay in token emission. By modifying $\lambda$, we encourage reduction in token emission delay.}
    \label{fig:appendix_seq_len_fastemit}
\end{figure}

In Section \ref{sec:durations_tdt}, we note that the large ratio of input sequence to target sequence $\frac{T}{U}$ was important to the emission of tokens with long durations, and that with a smaller ratio, the model would emit shorter duration tokens, even if it supported a large duration set ($N_d = 8$). In the following section, we attempt to analyze such a scenario, using different input sequence lengths ($T = \{ 20, 40, 70, 100 \}$) while maintaining the target sequence length of $U = 10$. This enables us to analyze the effect of modifying the ratio $\frac{T}{U}$. Note that as a result of changing $T$, $J_S$ is effectively a different sampled Joint tensor (represented as $J_{S_T}$), however, the target sequence $U$ remains the same.

%\vspace{-8pt}
In Fig.~\ref{fig:appendix_seq_len_non_fastemit}, we observe the alignments as we modify $T$. Of particular note is that when the ratio $\frac{T}{U}$ is small, tokens with short duration are preferred, performing nearly all token emissions without significant delay, and only towards the end does it emit a $d \in \{ 6, 8 \}$ duration token. On the other hand, when the ratio $\frac{T}{U}$ is large, tokens with long duration are preferred (with a large number of tokens having duration $d = 8$), and with a significant delay of token emission (note that FastEmit has not been enabled here). In Figure \ref{fig:appendix_seq_len_fastemit}, we see that increasing the FastEmit strength ($\lambda$) provides a corresponding decrease in token latency, even for large differences in sequence lengths.

\section{Robustness to Noise}
We measure the noise robustness of TDT models, by running inference on Librispeech test-clean augmented with noise samples in different signal-noise-ratios (SNRs). For each utterance, we randomly select a noise sample from MUSAN~\cite{snyder2015musan} and Freesound~\footnote{\url{https://freesound.org/}}. 
The noise sample is sub-segmented if it's longer than the utterance, or repeated if it's shorter than the utterance. The utterance samples are augmented with noise samples in 0, 5, 10, 15, and 20 SNRs.
We report the WER and inference time of conventional Transducers and TDT models with configuration 0-8. The results are shown in Figure~\ref{SNR}. As we can see, while Transducer and TDT models achieve very similar WERs in high SNR scenarios, as more noise is added, TDT models gradually outperform Transducers with larger and larger margins.  We also see that despite the SNR changes, the inference time for TDT only has minimal increases. This shows that TDT models have the capacity to perform much better in noisy conditions than conventional Transducers, in terms of both accuracy and speed aspects.

% \begin{table}[b]
%     \centering
%     \begin{tabular}{ccccc}
%     \toprule
%         & \multicolumn{2}{c}{Transducer}  & \multicolumn{2}{c}{TDT} \\
%        SNR          & WER   & time & WER & time \\
%     \midrule
%         $+\infty$   & 2.14 & 256 & 2.11 & 115 \\
%         20          & 2.40 & 248 & 2.36 & 117 \\
%         15          & 2.73  & 252 & 2.68 & 117 \\
%         10          & 3.71  & 251   & 3.52 & 117 \\
%         5           & 6.61 & 250 &  6.17 & 118 \\
%         0           & 14.24 & 248 & 12.85 & 118 \\
%         \bottomrule
        
%     \end{tabular}
%     \caption{Comparison of TDT VS Transducers on Librispeech test-clean with artificial noise at different SNRs. TDT model uses 0-8 configuration. The original test-clean dataset is listed as having SNR $+\infty$.}
%     \label{snr}
% \end{table}

\begin{figure}[t]
    \centering
    \includegraphics[scale=0.6]{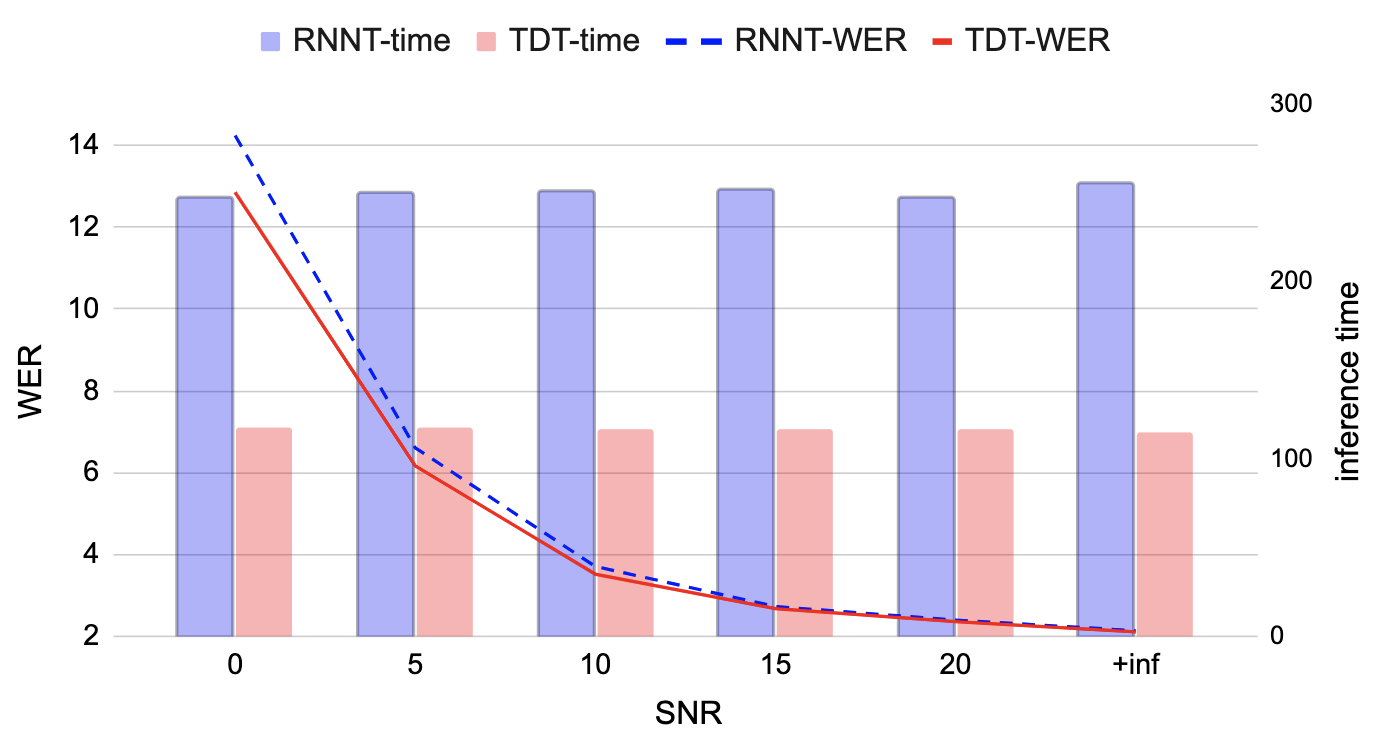}
    \caption{Comparison of TDT VS RNNT on Librispeech test-clean with noise added at different SNRs. TDT model uses 0-8 configuration. The original test-clean dataset is listed as having SNR = +inf. We see while TDT and RNNT achieve similar WER at low noise conditions, TDT is more robust to noise in low SNR settings. We also notice that the inference time stays constant for all TDT models, and this shows having additional noise in the audio does not change how the model emits long durations in inference. This Figure is the same as Figure \ref{SNR}, placed here for easy access.}
    \label{SNR2}
\end{figure}

%%%%%%%%%%%%%%%%%%%%%%%%%%%%%%%%%%%%%%%%%%%%%%%%%%%%%%%%%%%%%%%%%%%%%%%%%%%%%%%
%%%%%%%%%%%%%%%%%%%%%%%%%%%%%%%%%%%%%%%%%%%%%%%%%%%%%%%%%%%%%%%%%%%%%%%%%%%%%%%

\section{Robustness to Token Repetitions}
We notice that RNN-T models often suffer serious performance degradation when the text sequence has repetitions of the same tokens (repetition on the subword level to be exact if using subword tokenizations). Our investigation shows that more training data will not solve this issue, and this is an intrinsic issue of RNN-Ts.

Let's demonstrate the issue with an example here: suppose we have audio with text sequence \emph{two two two two two two two five}. Those words are frequent enough that they are all part of the BPE vocabulary. Let's assume in the audio, frames 0 to frame 40 correspond to all the \emph{two}s, and \emph{five} starts at frame 41; let's also assume we are at audio frame 30, and the model just emitted 5 \emph{two}s during  decoding. At this time, the decoder state was updated by feeding in 5 \emph{two}s.
%however, the decoder most likely does not know exactly how many \emph{two}'s are in the history. %\footnote{For stateless decoders, unless the history size is over 5, it is impossible for the decoder to know this; LSTM could know this ``in theory'', but in practice, they rarely learn to count the repetition numbers right.}. 
At this point, there are two possibilities,

% \begin{enumerate}
% \item 
\textbf{Option 1.} If the model emits another \emph{two} between frame 30 and 40, say at frame $t$, this means,
\begin{equation}
    \text{two} = \text{argmax} \big( \text{join}(\text{enc}[t], \text{dec(\textless bos\textgreater,} \underbrace{\text{two, two, two, two, two}}_\text{5 twos} ) \big)
\end{equation}
where join, enc, dec represent the computation of joiner, encoder, and decoder of the RNN-T model; for convenience, we assume the argmax operation directly returns the word from the distribution generated by the joiner. dec(a, b, c, d, ...) represents the final output of the decoder, after we sequentially pass a, b, c, d, ... as the decoder input. Since an LSTM decoder\footnote{Or in the case of stateless decoders, if the context size is less than 5, then it should be strictly equal.} rarely has memory beyond 3-4 words, we have 
\begin{equation}
\text{dec(\textless bos\textgreater}, \underbrace{\text{two, two, two, two, two}}_\text{5 twos}) \approx \text{dec(\textless bos\textgreater}, \underbrace{\text{two, two, two, two, two, two}}_\text{6 twos})
\end{equation}
We would like to point out that having a large number of repetitions isn't the necessary condition for this to happen; sometimes this happens with just two repetitions of the same token. 
Since \emph{two} is a non-blank emission, $t$ will not get incremented, and the next decoding step operates on the same enc($t$). 
Therefore, when we compute the output distribution of the next decoding step, it's likely that,
\begin{equation}
    \text{join}(\text{enc}[t], \text{dec(\textless bos\textgreater}, \underbrace{\text{two, two, two, two, two, two}}_\text{6 twos}) \approx \text{join}(\text{enc}[t], \text{dec(\textless bos\textgreater}, \underbrace{\text{two, two, two, two, two}}_\text{5 twos}) 
\end{equation}
Note, since joiner usually has a non-linearity in its computation, this does not strictly follow; although based on what we observed this is usually the case. The equations above are not meant to be rigorous proof but only serve to explain the issue.

Therefore in this next decoding step, it is likely that,
\begin{equation}
    \text{two} = \text{argmax} \big( \text{join}(\text{enc}[t], \text{dec(\textless bos\textgreater}, \underbrace{\text{two, two, two, two, two, two}}_\text{6 twos}) \big)
\end{equation}
i.e. the model emits \emph{two} for a second time at frame $t$. 
This will likely keep happening for 3 \emph{two}s, 4 \emph{two}s, etc, causing an infinite loop and won't terminate unless some \emph{max-symbol-per-decoding-step} is implemented in the decoding algorithm. In this case, we will end up having a lot of insertion errors in the output in the form of the same token repeating too many times.

\textbf{Option 2.} if the model keeps emitting all blanks until somewhere after frame 41, and then it  emits a \emph{five}. Then we would have deletion errors in the decoding output.
% \end{enumerate}

TDT is less prone to such repetition issues because the duration output of the model makes it not likely to stay on the same frame at different decoding steps (refer back to Fig.~\ref{fig:distribution}, there are very rare cases when duration 0 is emitted). Due to a lack of datasets specifically made with text repetitions, we use NeMo-TTS to generate 100 utterances, which randomly pick three digits from 1 to 9, and repeat each digit 3 - 5 times. We run ASR with different models on this dataset and results are  reported in Table \ref{repeated_wer2}. We see that TDT models are much more robust than RNN-Ts with repeated speech.

\begin{table}[h!]
    \centering
    \begin{tabular}{c c}
    \toprule
        model & WER\%\\
    \midrule
        RNNT-LSTM & 59.95 \\
        RNNT-stateless & 64.62 \\
        TDT [0-2] & 12.59 \\
        TDT [0-4] & 9.35 \\
        TDT [0-6] & 6.12 \\
        TDT [0-8] & 5.78 \\
    \bottomrule
    \end{tabular}
    \caption{WERs with different Transducer models on TTS generated dataset with repeated digits. This table is the same as Table \ref{repeated_wer} placed here for easy access.}
    \label{repeated_wer2}
\end{table}

\end{document}